\documentclass[11pt]{article}

\usepackage{amssymb}
\usepackage{mathrsfs}
\usepackage{enumitem}
\usepackage{tikz}
\usepackage{chet}

\allowdisplaybreaks

\usetikzlibrary{decorations.markings,decorations.pathmorphing,arrows}

\tikzset{cross/.style={path picture={
      \draw[black]
            (path picture bounding box.south east) --
            (path picture bounding box.north west)
            (path picture bounding box.south west) --
            (path picture bounding box.north east);}}}

\newcommand{\CH}{\mathcal{H}}
\newcommand{\CO}{\mathcal{O}}
\newcommand{\CI}{\mathcal{I}}
\newcommand{\dfour}{\ensuremath{d^{\hspace{0.5pt}4}\hspace{-0.5pt}}}
\newcommand{\dtwo}{\ensuremath{d^{\hspace{0.5pt}2}\hspace{-0.5pt}}}
\newcommand{\osbx}[2]{\ensuremath{x_{#1\hspace{1pt}}{\!}^{#2}}}

\DeclareMathOperator{\Tr}{Tr}
\DeclareMathOperator{\Disc}{Disc}
\DeclareMathOperator{\Ima}{Im}
\DeclareMathOperator{\Li}{Li}

\date{January 2014}

\title{OPE Methods for the Holomorphic Higgs Portal}

\author{Piyush Kumar, Daliang Li, David Poland, and Andreas
Stergiou\emails{(\href{mailto:piyush.kumar@yale.edu}{piyush.kumar},
\href{mailto:daliang.li@yale.edu}{daliang.li},
\href{mailto:david.poland@yale.edu}{david.poland},
\href{mailto:andreas.stergiou@yale.edu}{andreas.stergiou})@yale.edu}}

\affiliation{Department of Physics, Yale University, New Haven, CT 06511
USA}

\abstract{We develop a systematic and general approach to study the
effective Higgs Lagrangian in a supersymmetric framework in which the Higgs
fields in the visible sector couple weakly to another sector. The extra
sector may be strongly coupled in general. It is assumed to be
superconformal in the ultraviolet, but develop a mass-gap with
supersymmetry breaking in the infrared. The main technique used in our
approach is that of the operator product expansion (OPE). By using OPE
methods we are able to compute the parameters in the Higgs Lagrangian to
quadratic order and make general statements that are applicable to many
classes of models.  Not only does this approach allow us to understand the
traditional problems plaguing simple models from a different perspective,
it also reveals new possibilities for solutions of these problems. The
methods and results of our work should be useful in constructing a viable
and natural model of physics beyond the Standard Model.}

\begin{document}

\maketitle
\toc

\newsec{Introduction}
With the recent discovery of the Higgs boson at the LHC, it is safe to
assume that the electroweak symmetry is broken by the Higgs mechanism.
However, the observed mass of the Higgs boson near 126 GeV, and the null
evidence (so far) for any beyond-the-Standard-Model (BSM) physics, has
begun to pose important questions about the notion of ``electroweak
naturalness.''  For example, within the context of low-scale supersymmetry,
minimal realizations such as the (R-parity conserving) minimal
supersymmetric standard model (MSSM) are becoming increasingly fine-tuned
from an electroweak-scale point of view, at least in the technical sense of
't Hooft. One approach to this situation is to keep the BSM model minimal
(like the MSSM, for example) and accept some fine-tuning as a part of
nature~\cite{Wells:2004di, ArkaniHamed:2012gw, Arvanitaki:2012ps}.  An
alternative is to try to come up with not-so-minimal (but hopefully
well-motivated) models that are either fully electroweak-natural, or at
least alleviate the fine-tuning in the Higgs potential.

For the latter approach it is desirable to develop a unified treatment for
both perturbative and strongly-coupled models, both to guide the intuition
as well as for computational ease. In this work we take some steps in this
direction.  We consider a framework in which the Higgs fields in the MSSM
couple directly to another sector responsible for supersymmetry breaking
and its mediation, i.e.\ the messenger and/or supersymmetry-breaking
sector, via the superpotential
\eqn{W=\lambda_{u}\CH_{u}\CO_{u}+\lambda_{d}\CH_{d}\CO_{d}.}[W]
Here $\CO_{u}$ and $ \CO_{d}$ are in general composite operators that
belong to an SM representation conjugate to that of $\CH_u$ and $\CH_d$
respectively.  Models with such terms have been studied previously in the
literature in various contexts.  A well-studied situation is that $\CO_{u}$
and $\CO_{d}$ are composed of vector-like matter fields in a weakly-coupled
hidden sector~\cite{Dvali:1996cu, Martin:2009bg,
Martin:2010dc,Graham:2009gy,Craig:2012xp}. Examples of models where the
hidden sector is strongly-coupled include~\cite{Azatov:2011ht,
Azatov:2011ps,Gherghetta:2011na, Kitano:2012wv,Evans:2012uf}. The presence
of additional sectors with these couplings is also well-motivated from a
top-down (string theory) point of view, as in~\cite{Heckman:2011bb,
Heckman:2012nt, Heckman:2012jm}. Finally, such scenarios can be studied
very generally by expressing observable parameters in terms of hidden
sector correlators~\cite{Meade:2008wd,Komargodski:2008ax,Craig:2013wga}.
It is clear that the range of hidden-sector models included in \W can run
the gamut from weakly- to strongly-coupled. In general, the dynamics of
such a sector and its coupling to the Higgs can have important effects on
various terms in the Higgs potential as well as Higgs couplings.  For
example, the quadratic terms in the Higgs potential, which are determined
by $\mu$ and the supersymmetry-breaking parameters in the Higgs sector, as
well as the quartic terms, which determine the physical Higgs boson masses
and mixings, are affected in general.

In this paper we focus on the computation of the quadratic terms in the
Higgs potential. The emphasis of our work is to develop techniques that are
applicable to a large class of models, even those in which the additional
sector is strongly coupled. The primary tool that we will use in this
regard is the operator product expansion (OPE).  On general grounds, since
local physics should be captured by local operators, it is expected that
the product of two nearby operators can be replaced by a linear combination
of local operators,
\eqn{\CO_i(x)\CO_j(0) = \sum_{k} c_{\smash{ij}}^{\phantom{ij\!}k}(x)
\CO_k (0),}[OPE]
with $c_{\smash{ij}}^{\phantom{ij\!}k}$ referred to as OPE or Wilson
coefficients.  All $x$-dependence of the operator product
$\CO_i(x)\CO_j(0)$ is included in $c_{\smash{ij}}^{\phantom{ij\!}k}$.  In
conformal theories \OPE is a convergent expansion inside correlation
functions where no other operators are within a distance $x$ from the
origin, while in in nonconformal theories \OPE is in general an asymptotic
expansion valid in the limit $x \rightarrow 0$. In such cases the OPE is a
very powerful way of separating high- from low-energy physics, for the
Wilson coefficients $c_{\smash{ij}}^{\phantom{ij\!}k}$ are determined by UV
physics, while the expectation values of operators on the right-hand side
are determined by IR physics. Therefore, if the UV physics is under
control, the OPE can be used to gain an understanding of the dynamics even
in the IR.

One possibility is that the UV physics is asymptotically free, as in the
case of QCD, where OPE techniques have been extensively used to determine
important constraints on the low energy theory, via QCD sum
rules~\cite{Shifman:1978bx}. OPE methods and dispersion relations have also
been used in computing the cross-section of $e^+e^- \rightarrow
\text{hadrons}$. There, the electromagnetic current-current correlation
function $\langle J_{EM}(x)J_{EM}(0)\rangle$ can be computed by replacing
it with the OPE and then using dispersion relations to obtain it in the
physical region.

Another possibility is that the UV physics for the additional sector has a
large symmetry group, such as the superconformal group.  Of course, the IR
physics is neither conformal (the states in the additional sector
ultimately acquire a mass) nor supersymmetric (the multiplets in the
additional sector have mass splittings $\propto \sqrt{F}$ if they couple to
supersymmetry breaking).  But it is still possible to apply OPE techniques
if the symmetries are regarded as spontaneously broken, since then they are
restored in the UV. Just like in QCD, then, OPE methods can be used to
describe low-energy observables. This was demonstrated
in~\cite{Fortin:2011ad}, utilizing results
of~\cite{Meade:2008wd,Buican:2008ws,Fortin:2011nq}, for the SM gauge
current-current correlation functions $\langle J_{a}(x)J_{a}(0)\rangle$,
$a=1,2,3$, from which gaugino and sfermion masses could be obtained.

In this work we will use OPE techniques to compute parameters in the Higgs
Lagrangian when the Higgs fields couple to an additional sector via \W, and
in which the additional sector has (at least an approximate) superconformal
symmetry in the UV.  Starting from the expressions
of~\cite{Komargodski:2008ax}, where soft parameters are expressed in terms
of two-point functions of hidden-sector operators, we will show, utilizing
the general formalism of~\cite{Osborn:1998qu}, that the approximate
superconformal symmetry provides powerful constraints on the form of such
terms. Our strategy is to use the (kinematically constrained) form of
three-point functions in $\mathcal{N}=1$ superconformal theories to extract
the OPE of the first two operators with the third. In doing that, we can
identify classes of hidden-sector operators that contribute to the two-point
functions of~\cite{Komargodski:2008ax}, and consequently the parameters in
the Higgs Lagrangian.

Our treatment brings a new perspective on the $\mu/B_\mu$
problem~\cite{Dvali:1996cu} and the $A/m_H^2$ problem~\cite{Craig:2012xp}
in models of gauge-mediated supersymmetry breaking, and also opens up new
possibilities for viable electroweak symmetry breaking (EWSB). As a
nontrivial consistency check of our methods we reproduce the well-known
results for the Higgs soft terms in the weakly-coupled toy model
of~\cite{Dvali:1996cu}. Finally, we make some comments on the computation
of the quartic terms in the Higgs potential via OPE techniques, but a
detailed computation is left for the future.

The paper is organized as follows. In section \ref{setup} we lay out our
setup and describe our assumptions in detail, followed by a brief
description of the OPE formalism and the constraints from superconformal
symmetry. Section \ref{computations} forms the technical meat of the paper,
in which the computations of the soft terms and $\mu$ by the OPE method are
outlined. In section \ref{toy} we reproduce the results for the weakly
coupled model of~\cite{Dvali:1996cu} for the Higgs soft terms using the
OPE, and we elucidate some subtle features of perturbative computations
with the OPE. Section \ref{pheno} is devoted to a broad discussion of
various phenomenological implications of the results obtained.  Finally, in
section \ref{discuss} we summarize our main results, and make comments
about future directions. Some technical details of computations regarding
the projection of the two-point function of $\mathcal{N}=1$ superconformal
primary operators to two-point functions of their conformal primary
components are given in Appendix \ref{technicalTwoP}. In Appendix
\ref{technicalThreeP} we give details on the projection of superconformal
three-point functions to conformal primary ones, and we construct the
associated OPEs of conformal primary operators.

\newsec{The Higgs effective Lagrangian}[setup]
In this section we describe the setup in some detail. We are interested in
a framework in which the Higgs fields in the MSSM couple to operators
$\CO_u$ and $\CO_d$ in another sector via the superpotential
\eqn{W=\lambda_{u}\CH_{u}\CO_{u}+\lambda_{d}\CH_{d}\CO_{d}=
\lambda_u\epsilon^{ij}(\CH_u)_i(\CO_u)_j+
\lambda_d\epsilon_{ij}(\CH_d)^i(\CO_d)^j.}[]
Here $\CO_u$ and $\CO_d$ are $SU(2)$ doublet operators of dimensions
$\Delta_{\CO_u}$ and $\Delta_{\CO_d}$, with hypercharges opposite to that
of $\CH_u$ and $\CH_d$ respectively ($i,j$ are $SU(2)$ indices). The
couplings $\lambda_u$ and $\lambda_d$ are assumed to be perturbative at the
mass scale $M$ characterizing the additional sector.\footnote{Concretely,
the dimensionless renormalized couplings $\lambda_i(\mu) \equiv \lambda_i
\mu^{2-\Delta_{\CO_i}}$ should satisfy $\lambda_i(M) \ll 1$.} Note that
this still allows the sector in which $\CO_u$ and $\CO_d$ belong to be
strongly coupled.  We will consider the situation in which $\CO_u$ and
$\CO_d$ belong to a sector responsible for supersymmetry breaking and its
mediation to the visible sector. As such, these operators could just be
part of the supersymmetry-breaking sector, or they could comprise a
``messenger'' sector distinct from the supersymmetry-breaking sector but
coupled to it by a weak coupling (say $\kappa$). The latter case gives rise
to a framework which has been called ``general messenger Higgs mediation
(GMHM)'' in~\cite{Craig:2013wga}. In this case, a double expansion in
$\lambda_{u,d}$ and $\kappa$ is possible.  In this work, however, we will
focus on the more general case where only an expansion in $\lambda_{u,d}$
is available, although we will make some comments about the more special
case with a coupling $\kappa$.

Now, by integrating out the dynamics of the sector containing $\CO_u$ and
$\CO_d$, one generates various terms in the Higgs Lagrangian. The terms in
the Higgs Lagrangian at quadratic order and zero momentum are
\eqna{\mathscr{L} &= \mathcal{Z}_uF_{H_u}^{\dag}F_{H_u} +
\mathcal{Z}_dF_{H_d}^{\dag}F_{H_d}+\left(\mu \int \dtwo\theta\,
\CH_u\CH_d+ \text{c.c.}\right)- V^{(\text{soft})}_{\text{Higgs}}
-V^{(\text{other})}_{\text{Higgs}},\\
V^{(\text{soft})}_{\text{Higgs}} &=
m_{H_u}^2H_u^{\dag}H_u + m_{H_d}^2H_d^{\dag}H_d + (B_\mu H_uH_d
+ \text{c.c.)} + (A_uH_uF_{\CH_u}^{\dag}+A_dH_dF_{\CH_d}^{\dag}
+\text{c.c.}),\\
V^{(\text{other})}_{\text{Higgs}} &= (a'_uH_u F_{\CH_d}
+\text{c.c.})
+ (a'_dF_{\CH_u} H_d+\text{c.c.})+(\gamma F_{\CH_u}F_{\CH_d}
+\text{c.c.}).}[LHiggs]
Some comments are in order. Here we have assumed that there is no bare
$\mu$ term in the superpotential---it is only generated after integrating
out the additional sector. Also, in addition to the usual soft
supersymmetry breaking terms in the Higgs potential
$V^{(\text{soft})}_{\text{Higgs}}$, there are additional terms, which we
collectively denote as $V^{(\text{other})}_{\text{Higgs}}$.

In a supersymmetry-breaking vacuum in a globally supersymmetric theory (as
considered here), the vacuum energy is strictly positive. In the typical
case where some operator in the additional sector breaks supersymmetry with
its F-term vacuum expectation value (vev), $F$, it can be shown that the
vacuum energy density is equal to $|F|^2$.  If $\sqrt{F}$ is smaller than
the typical mass scale $M$ of operators in the additional sector, then all
observables can be expanded in powers of $F/M^2$. In many models the terms
in $V^{(\text{other})}_{\text{Higgs}}$ are typically suppressed by powers
of $F/M^2$ relative to the terms in $V_{\text{Higgs}}^{(\text{soft})}$ and
the $\mu$ parameter, and can thus be neglected. We will show, however,
that, at least in principle, some of the terms in
$V^{(\text{other})}_{\text{Higgs}}$ may not be suppressed relative to the
$\mu$ parameter. We will also explain why such suppressions are so common
in models found in the literature.

The parameters appearing in the Higgs Lagrangian can be computed in terms
of the zero-momentum limit of two-point correlation functions involving
components of $\CO_u$ and $\CO_d$. This is easy to see from an
effective-field theory point of view, and can be explicitly worked out by
expanding $\exp[i\int\!\dfour x\,(\int\!\dtwo\theta\,(\lambda_u\CH_u\CO_u
+\lambda_d\CH_d\CO_d)+\text{c.c.})]$ to quadratic order and matching with
the effective Lagrangian \LHiggs. This has already been done
in~\cite{Komargodski:2008ax}. The soft parameters and the $\mu$ term
generated at the scale $M$ due to the superpotential (\ref{W}) at leading
order in $\lambda_{u,d}$ are given by\footnote{We are following the
conventions of Wess \& Bagger~\cite{Wess:1992cp}, taking care of factors of
$i$ and minus signs as explained in the appendix of~\cite{Fortin:2011nq}.
The action of operators is always the adjoint action, i.e.\
$Q(O)\equiv[Q,O\}$.  Note that there is an extra factor of $\tfrac12$
compared to~\cite{Komargodski:2008ax} from taking into account the $SU(2)$
gauge indices.}
\eqna{\mu&=\frac{i}{8}\lambda_{u}\lambda_{d}\left.\left\langle\,
\int\dfour x\,e^{-i p\cdot x}Q^\alpha(O_{u}(x))Q_{\alpha}(O_{d}(0))
\right\rangle\right|_{p \rightarrow 0},\\
B_{\mu}&=\frac{i}{2^5}\lambda_{u}\lambda_{d}\left.\left\langle\,
\int\dfour x\,e^{-i p\cdot x}Q^2(O_{u}(x))Q^2(O_{d}(0))
\right\rangle\right|_{p \rightarrow 0},\\
\delta A_{u,d}&=-\frac{i}{8}|\lambda_{u,d}|^{2}\left.\left\langle\int\dfour
x\,e^{-i p\cdot x}Q^{2}(O_{u,d}(x)O_{u,d}^{\dagger}(0))\right\rangle
\right|_{p \rightarrow 0},\\
\delta m_{H_{u,d}}^{2}&=-\frac{i}{2^5}|\lambda_{u,d}|^{2}\left.\left\langle
\int\dfour x\,e^{-i p\cdot x}Q^2\bar{Q}^{2}(O_{u,d}(x)O_{u,d}^
{\dagger}(0))\right\rangle\right|_{p \rightarrow 0}.}[HiggsParams]
We will consider the contributions solely generated from the superpotential
$(\ref{W})$. Clearly, $\mu$ and $B_\mu$ arise from a \emph{chiral-chiral}
two-point function while $\delta A_{u,d}$ and $\delta m_{H_{u,d}}^2$  arise
from a \emph{chiral-antichiral} one.\foot{Note that the $\mu$-term in
(\ref{LHiggs}) actually gives rise to three terms in the
Lagrangian---$\mu\,(-\psi_{\CH_u}\psi_{\CH_d}+H_uF_{\CH_d}+
F_{\CH_u}H_d)+\text{c.c}$.} Since
\eqn{Q^2(Q^\alpha(O_u(x))Q_\alpha(O_d(0)))=-
Q^2(O_u(x))Q^2(O_d(0)),}[]
we see that there is a relation between $\mu$ and $B_\mu$, similar to the
relation between $\delta A_{u,d}$ and $\delta m_{H_{u,d}}^2$. Finally, note
that $B_\mu$, $\delta A_{u,d}$, and $\delta m_{H_{u,d}}^2$ can be written
as $\langle Q(\cdot) \rangle $, while $\mu$ cannot. Thus, the soft
parameters are generated only when supersymmetry is broken (as they must),
i.e.\ when $\langle 0| Q \neq 0$, while $\mu$ may receive contributions
consistent with supersymmetry.

Similarly, the terms in $V^{(\text{other})}_{\text{Higgs}}$ are given by
\eqna{a'_{u,d} &=\frac{i}{8}\lambda_{u}\lambda_{d} \left.\left\langle\,
\int\dfour x\,e^{-i p\cdot x}Q^\alpha(O_{u,d}(x)
Q_\alpha(O_{d,u}(0)))\right\rangle\right|_{p \rightarrow 0},\\
\gamma &=\frac{i}{2}\lambda_{u}\lambda_{d} \left.\left\langle\,
\int\dfour x\,e^{-i p\cdot x}O_{u}(x)O_{d}(0)\right\rangle
\right|_{p \rightarrow 0},\\
\mathcal{\delta Z}_{u,d} &=\frac{i}{2}|\lambda_{u,d}|^2\left.\left
\langle\,\int\dfour x\,e^{-i p\cdot x}O_{u,d}(x)O_{u,d}^{\dag}(0)
\right\rangle\right|_{p \rightarrow 0}.}[HiggsHard]
Again, $a'_{u,d}$ and $\gamma$ arise from a \emph{chiral-chiral}
correlation function while $\delta \mathcal{Z}_{u,d}$ arises from a
\emph{chiral-antichiral} one. As explained in~\cite{Komargodski:2008ax},
$\delta \mathcal{Z}_{u,d}$ corresponds to the contribution to the
(supersymmetric) wave-function renormalization of the Higgs fields due to
(\ref{W}), and affects the overall normalization of the physical
observables. In terms of the expansion in $\lambda_{u,d}$, these
corrections are negligible; hence we will not study $\delta
\mathcal{Z}_{u,d}$ in detail. On the other hand, it can be seen from
(\ref{LHiggs}) and (\ref{HiggsHard}) that $a'_{u,d}$ arises from effects
which cannot be captured by the $\mu$ term in the superpotential alone.
Finally, $\gamma$ gives rise to corrections to Higgs parameters as well as
interactions between four MSSM sfermions (after eliminating $F_{H_u}$ and
$F_{H_d}$). We will show using our methods that $\gamma$ must be suppressed
in phenomenologically viable models, but $a'_{u,d}$ may be generated at the
same order as $\mu$ in general.

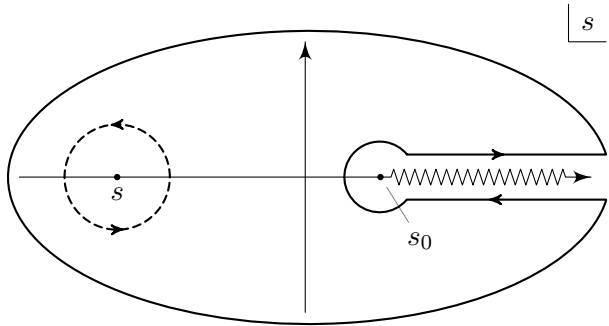
\begin{figure}[ht]
  \centering
  \begin{tikzpicture}[line join=round,line cap=round]
    \draw (-3.8,0)--(1,0);
    \draw[decorate,decoration={zigzag,segment length=4pt, pre
    length=4pt, post length=3pt, amplitude=3pt}] (1,0)--(3.6,0);
    \draw[decoration={markings,mark=at position 1 with
    {\arrow[scale=1.75,>=latex']{>}}}, postaction={decorate}]
    (3.6,0)--(3.8,0);
    \draw[decoration={markings,mark=at position 1 with
    {\arrow[scale=1.75,>=latex']{>}}}, postaction={decorate}]
    (0,-1.8)--(0,1.8);
    \draw (3.5,2.3)--(3.5,1.8)--(4,1.8) node[above left=0.5pt] {$s$};
    \node at (1,0) [pin=290:$s_0$] {};
    \filldraw[black] (1,0) circle (1pt);
    \filldraw[black] (-2.5,0) circle (1pt) node[below] {$s$};
    \draw[densely dashed,thick,decoration={markings,
    mark=at position 0.27 with{\arrow[thick,>=stealth',rotate=-5]{>}};,
    mark=at position 0.77 with {\arrow[thick,>=stealth',rotate=-5]{>}}},
    postaction={decorate}] (-2.5,0) circle (0.7cm);
    \draw[thick,decoration={markings,
    mark=at position 1.3 cm with {\arrow[thick,>=stealth']{>}};,
    mark=at position 22.84 cm with {\arrow[thick,>=stealth']{>}}},
    postaction={decorate}]
    (1.35,0.3)--(4,0.3) arc (9:351:4cm and 1.95cm)
    (4,-0.3)--(1.35,-0.3) arc (320:40:0.47cm);
  \end{tikzpicture}
  \caption{The dashed contour can be deformed to the solid contour giving
  \integrate.}
  \label{fig:DispRel}
\end{figure}

\subsec{OPE methods and approximations}[OPEmethod]
Given \HiggsParams and \HiggsHard, one would like to compute, at least
approximately, the relevant (momentum-space) correlation functions in a
manner applicable to many cases. The OPE is very useful in this regard: we
can apply the OPE to the two-point functions in \HiggsParams and
\HiggsHard, and identify operators that can obtain nonzero vevs consistent
with Poincar\'e and gauge invariance.\footnote{From the general properties
of any CFT, such operators $\CO$ must be scalar conformal primaries, i.e.\
such that $K_{\mu}(\CO(0)) = 0$. Such operators can only appear as the
descendants of superconformal primaries with Lorentz representation
$(j,\bar{\jmath})=\{(0,0), (0,\frac12), (\frac12,0), (\frac12,
\frac12)\}$.}  Of course, we also need to compute the necessary OPE
coefficients.  Those are computable in the regime of large space-like or
Euclidean (unphysical) momenta, but the answer can be straightforwardly
analytically continued to the physical region (large time-like momenta).
The critical observation, then, is that if the two-point function $A(s)$
(where $s=-p^2$) has a branch cut starting at some threshold $s_0$ and no
other singularities in the physical sheet, we can use the dispersion
relation (see Fig.~\ref{fig:DispRel})
\eqn{A(s)=\frac{1}{2\pi i} \int _{s_0}^\infty ds'\,
\frac{\Disc A(s')}{ s'-s}=\frac{1}{\pi}\int _{s_0}^\infty
ds'\, \frac{\Ima A(s')}{s'-s}.}[integrate]
To summarize, our strategy is the following: we approximate $A(s)$ by going
to large $s$, applying the OPE, keeping only the first few terms in the
$1/s$ expansion, and finally using the dispersion relation \integrate to
obtain an approximation to $A(s)$ even at $s=0$, as needed in \HiggsParams
and \HiggsHard.

Note that, beyond the truncation of the OPE (which can be avoided in some
tractable cases), there are two more approximations being used here. First,
although the OPE is strictly valid for large $s$, we apply it in the entire
region from $s_0$ to $\infty$. In our case we take $s_0 = (2M)^2$, where,
as mentioned earlier, $M$ is the typical mass scale in the additional
sector.  This is a good approximation as long as $M$ is at least somewhat
higher than the electroweak scale. The other approximation comes about when
we neglect the fact that the position $s_0$ of the branch point technically
depends on the masses of the states in the additional sector, which are
\emph{not} all at $M$ due to the presence of supersymmetry breaking.
However, this is a good approximation at least at small $F/M^2$, since the
OPE is insensitive to the precise positions of these branch points.

As we already mentioned, in manipulations leading to \integrate we assume
that there are no poles in $A(s)$ below the threshold $s_0$. This is a good
approximation in models where the operators in $A(s)$ belong to some
weakly-coupled sector, or perhaps certain classes of strongly coupled
sectors. Nevertheless, possible poles, either from bound states or
fundamental one-particle states, can make \integrate inaccurate.  This is
because the Wilson coefficients in the OPE used in the evaluation of $A(s)$
will get extra contributions from poles, beyond those captured by the
branch cuts in \integrate. While it would be interesting to do a more
general analysis including possible contributions from such poles, in this
work we will focus on the simplest analytical structure, describing
contributions to soft parameters which should be present in any model with
a threshold $s_0$. Moreover, it is important to stress that although the
computation of the Wilson coefficients is affected by the analytic
structure of the two-point function, the form of the OPE is completely
general and is not affected by the possible presence of poles, i.e.\ poles
do not result in extra operators in the right-hand side of the OPE.

\newsec{Computations of Higgs parameters}\label{computations}
As mentioned above, we are interested in computing the chiral-chiral OPE of
superfields $\CO_u$ and $\CO_d$, and the chiral-antichiral OPE of
superfields $\CO_{u,d}$ and $\bar{\CO}_{u,d}$. These OPEs can be written as
a sum over superconformal primary operators (those annihilated, at the
origin, by the $S^{\alpha}$ and $\bar{S}^{\dot{\alpha}}$ generators of the
superconformal algebra) and their descendants (obtained by acting with
$Q_{\alpha}$ and $\bar{Q}_{\dot{\alpha}}$). The OPE $\Phi_1\times \Phi_2$
of two superconformal primary superfields $\Phi_1$ and $\Phi_2$ will
contain a superconformal primary superfield $\CO$ only if the superspace
three-point function $\langle \Phi_1 \Phi_2 \CO^{\dag}\rangle$ is
nonvanishing. Therefore, for the case at hand, one has to study three-point
functions of the type $\langle \CO_u \CO_d \bar{\CO}^{\bar I}\rangle$ and
$\langle \CO_{u,d} \bar{\CO}_{u,d} \bar{\CO}^{\bar I} \rangle$, where
$\CO^I$ is a general superconformal primary superfield with Lorentz index
$I$. The Lorentz index $I$ can be labeled by spins $(j, \bar{\jmath})$
according to the representation of $SO(4) \simeq SU(2) \times SU(2)$. It is
also customary to label the spin for traceless symmetric tensor
representations by $j = \bar{\jmath} = \frac{\ell}{2}$.

\subsec{Superconformal two-point functions}
In an $\mathcal{N}=1$ superconformal theory the two-point function of
superconformal primary operators is determined up to a constant. Moreover,
operators of different scaling dimensions have vanishing two-point
functions with each other.  The two-point function of an operator $\CO^I$
with its Hermitian conjugate ${\bar{\CO}}^{\bar{I}}$ (bars and daggers are
both used in this work to denote Hermitian conjugation) is given by
\eqn{\langle\CO^I(z_1){\bar{\CO}}^{\bar{I}}(z_2)\rangle =
C_\CO\frac{\CI^{I\bar{I}}(x_{1\bar{2}},x_{\bar{1}2})}
{\osbx{\bar{1}2}{2\bar{q}}\osbx{\bar{2}1}{2q}},}[TwoPoints]
where $C_\CO$ is positive coefficient\foot{A basis of operators can always
be chosen so that all $C_\CO$'s are equal to one, but in this work we will
keep the $C_\CO$'s explicit.} in a unitary theory, and $\CI^{I\bar{I}}$ is
an appropriate tensor that accounts for the Lorentz structure of the
left-hand side.  For example, for a vector operator $\CO^\mu$ we can take
$\CI^{\mu\nu}= \Tr(\bar{\sigma}^\mu\text{x}_{1\bar{2}}
\bar{\sigma}^\nu\text{x}_{2\bar{1}})/2\sqrt{\osbx{\bar{1}2}{2}
\osbx{\bar{2}1}{2}}.$ The charges $q$ and $\bar{q}$ are such that the
scaling dimension and the R-charge of $\CO^I$ are given by
$\Delta=q+\bar{q}$ and $R=\tfrac23(q-\bar{q})$ respectively. For more
details the reader is referred to~\cite{Osborn:1998qu}.

\subsec{Three-point functions for the chiral-chiral OPE}[chiralchiral]
A formalism for describing the three-point functions of $\mathcal{N}=1$
superconformal primary operators was presented in~\cite{Osborn:1998qu} in
full generality. The case of two scalar chiral superfields with a general
third superfield was worked out in~\cite{Poland:2010wg,Vichi:2011ux}. The
general form of the three-point function in this case is
\eqn{\langle \Phi_{1}(z_{1+})\Phi_{2}(z_{2+})\bar{\CO}^{\bar{I}}(z_{3})
\rangle=\frac{\lambda_{12\CO}}{\osbx{\bar{3}1}{2\Delta_{1}}
\osbx{\bar{3}2}{2\Delta_{2}}}{\bar{t}}^{\bar{I}}(\bar{X}_{3},\Theta_{3},
\bar{\Theta}_{3}),}[ChiralChiralThreePF]
where $\lambda_{12\CO}$ is a complex coefficient in general, $z$ represents
general superspace coordinates \{$x, \theta,\bar{\theta}$\}, and $z_{+}$
represents the chiral superspace coordinates \{$y, \theta$\}, with
$y=x+i\theta \sigma \bar{\theta}$. The supersymmetric interval between
points $x_i$ and $x_j$ is given by
\eqn{x_{\bar{\imath}j}= - x_{j\bar{\imath}}\equiv \bar{y}_{i}-y_{j}+
2i\theta_{j}\sigma\bar{\theta}_{i}=x_{ij}-i\theta_i\sigma\bar{\theta}_i
-i\theta_j\sigma\bar{\theta}_j+2i\theta_j\sigma\bar{\theta}_i,}[]
and so
\eqn{\osbx{\bar{\imath}j}{2}=\osbx{ij}{2}-2i\theta_i
x_{ij}\cdot\sigma\bar{\theta}_i -2i\theta_jx_{ij}\cdot\sigma\bar{\theta}_j
+4i\theta_jx_{ij}\cdot\sigma\bar{\theta}_i +2\theta_i^2
\bar{\theta}_i^{\hspace{1pt}2}+2\theta_j^2\bar{\theta}_j^{\hspace{1pt}2} -8\theta_i\theta_j\bar{\theta}_i^{\hspace{1pt}2}
-8\theta_j^2\bar{\theta}_i\bar{\theta}_j
+8\theta_j^2\bar{\theta}_i^{\hspace{1pt}2},}[]
where $x_{ij}=x_i-x_j$.  The quantities $\Theta_3, \bar{\Theta}_3$ and
$X_3,\bar{X_3}$ are given by
\eqn{\bar{\Theta}_{3}=i\left(\frac{1}{\osbx{\bar{3}1}{2}}\theta_{31}
\text{x}_{\smash{1\bar{3}}}-\frac{1}{\osbx{\bar{3}2}{2}}\theta_{32}
\text{x}_{\smash{2\bar{3}}}\right)=\Theta_3^\dagger, \qquad
\bar{X}_{3}^{\mu}=\frac12\frac{x_{3\bar{2}\nu}x_{\bar{2}1\rho}
x_{1\bar{3}\sigma}}{\osbx{\bar{2}3}{2}\osbx{\bar{3}1}{2}}
\Tr(\bar{\sigma}^{\mu}\sigma^{\nu}\bar{\sigma}^{\rho}\sigma^{\sigma})=
(X_3^\mu)^\dagger.}[]
In the following we define upright quantities $\text{X}$ by exchanging a
Lorentz vector index with a pair of dotted-undotted spinor indices with the
use of the Pauli matrices, i.e.\ $\text{X}_{\alpha\dot{\alpha}}=
\sigma^\mu_{\alpha\dot{\alpha}} X_\mu.$ For example,
\eqn{\bar{\text{X}}_3=-\frac{\text{x}_{3\bar{2}}
\text{x}_{\bar{2}1} \text{x}_{1\bar{3}}}{\osbx{\bar{2}3}{2}
\osbx{\bar{3}1}{2}}=\text{X}_3^\dagger.}[]

The chirality of $\Phi_{1,2}$ implies that the function $\bar{t}^{\bar{I}}$
can only depend on $\bar{X}_3$, $\bar{\Theta}_3$ and must satisfy a Ward
identity.  These constraints are obeyed by the three families of solutions
listed below.
\begin{enumerate}[ref=Solution~(\Roman{enumi}),align=left,leftmargin=*,
    labelindent=\parindent]
  \renewcommand{\labelenumi}{Solution~(\Roman{enumi}):}
  \item\label{SolOne}
    \eqn{{\bar{t}}_{1}(\bar{X}_{3},\bar{\Theta}_{3})=1,\qquad
    (\Delta_{\CO}=\Delta_{1}+\Delta_{2},\, R_{\CO}=R_{1}+R_{2}),}[]
  \item\label{SolTwo}
    \eqn{{\bar{t}}_{2\,\alpha_2\ldots\alpha_\ell}^{\dot{\alpha}_1\ldots
    \dot{\alpha}_\ell}(\bar{X}_{3},\bar{\Theta}_{3})=
    \bar{\Theta}_{3}^{(\dot{\alpha}_{1}}
    \bar{\text{X}}_{3\alpha_{2}}^{\dot{\alpha}_{2}}\cdots
    \bar{\text{X}}_{3\alpha_{\ell}}^{\dot{\alpha}_{\ell})},
    \qquad(\Delta_{\CO}=\Delta_{1}+\Delta_{2}+\ell-\tfrac12,\,
    R_{\CO}=R_{1}+R_{2}-1),}[]
  \item\label{SolThree} \eqn{{\bar{t}}_{3}^{\mu_1\ldots\mu_\ell}
    (\bar{X}_{3},\bar{\Theta}_{3})=
    \bar{\Theta}_{3}^{2}\frac{\bar{X}_{3}^{\mu_{1}}\cdots
    \bar{X}_{3}^{\mu_{\ell}}}{\bar{X}_{3}^{\Delta_{1}+\Delta_{2}
    -\Delta_{\CO}+\ell+1}},\qquad
    (\Delta_{\CO}\ge|\Delta_{1}+\Delta_{2}-3|+\ell+2,\,
    R_{\CO}=R_{1}+R_{2}-2).}[]
\end{enumerate}
Here $\Delta_{\CO}$ and $R_{\CO}$ are the scaling dimension and
superconformal R-charge of the operator $\CO$ respectively (similar
notation for others). We will denote the operators for solution $n$ and
spin $\ell$ by $\CO_{n,\ell}$.

Among the various operators in the OPE of the component fields of
$\Phi_{1,2}$ we are only interested in scalar conformal primaries
(annihilated by $K_{\mu}$ at the origin) that could develop nonzero vevs
without breaking Poincar\'e invariance and SM gauge invariance.  Note that
such scalar conformal primaries can either be superconformal primaries or
superconformal descendants. Now, in order to contain a scalar component, an
$\mathcal{N}=1$ superfield must have $j+\bar{\jmath} =
0,\,\frac{1}{2},\,1$.  Thus, the corresponding three-point functions must
be given by the $\ell=0$ case of \ref{SolOne}, the $\ell=1$ case of
\ref{SolTwo}, or the $\ell=0,1$ cases of \ref{SolThree}. In the notation
described above, the contributing operators are then $\CO_{1,0}$,
$\CO_{2,1}^{\alpha}$, $\CO_{3,0}$, and $\CO_{3,1}^{\mu}$, which will have
component expansions of the form
\eqn{\begin{gathered}
\mathcal{O}=O+i\theta Q O + i\bar{\theta}\bar{Q}O
+\tfrac{1}{4}\theta^{2}Q^2O+\tfrac14\bar{\theta}^{\hspace{1pt}2}
\bar{Q}^2O+\cdots,\\
\mathcal{O}_{\alpha}=O_{\alpha}-\tfrac{i}{2}\theta_{\alpha}Q^{\beta}
O_{\beta}+\cdots,\\
\mathcal{O}^{\mu}=O^{\mu}+i\theta QO^{\mu}+i\bar{\theta}\bar{Q}O^{\mu}+
\tfrac{1}{8}\theta\sigma^{\mu}\bar{\theta}(Q\sigma^{\nu}\bar{Q}O_{\nu})
+\cdots.
\end{gathered}}[]

Note that in the $\theta$-expansion of a superfield one gets component
fields that are not necessarily conformal primaries. As explained in
Appendix~\ref{technicalTwoP}, this happens at order $\theta\bar{\theta}$
and higher. In this work, when it is necessary, we will indicate that an
operator is conformal primary by writing it as $[\,\cdot\,]_p$.

\subsec{Three point functions for the chiral-antichiral OPE}
The superconformal three-point function of a chiral superfield, an
antichiral superfield and a general third operator was worked out
in~\cite{Poland:2010wg}, with the result
\eqn{\langle
\Phi(z_{1+})\bar{\Phi}(z_{2-}){\bar{\CO}}^{\mu_{1}\ldots\mu_{\ell}}
(z_3)\rangle=\frac{\lambda_{1\bar{2}\CO^{\mu_1\ldots\mu_\ell}}}
{\osbx{\bar{2}3}{2\Delta_{\Phi}}\osbx{\bar{3}1}{2\Delta_{\Phi}}}
\bar{X}_{3}^{\Delta_{\CO}-2\Delta_{\Phi}-\ell}\bar{X}_{3}^{\mu_{1}}\cdots\bar{X}_{3}^{\mu_{\ell}},
}[ChiralAntichiralThreePF]
where
\eqn{\bar{X}_{3}^{2}=\frac{\osbx{\bar{2}1}{2}}{\osbx{\bar{2}3}{2}
\osbx{\bar{3}1}{2}}.}[]
As opposed to the chiral-chiral case, where there were three classes of
structures in the right-hand side of \ChiralChiralThreePF, we see here that
the solution of the superconformal constraints results in a unique class of
structures for the chiral-antichiral three-point function.

\subsection{Results}
Armed with the above superfield three-point functions, we can expand these
in $\theta,\bar{\theta}$ to compute component three-point functions. Using
these, we can then extract the terms in the right-hand side of the relevant
chiral-chiral and chiral-antichiral OPEs and compute contributions to the
parameters in the Higgs Lagrangian. We do this below, starting with
parameters which are determined by the chiral-chiral OPE. For details on
the derivation of the various expressions below the reader is referred to
the appendices.

\subsubsection{\texorpdfstring{$\mu$}{mu}}
From \HiggsParams, we see that $\mu$ is determined by $\langle
Q^{\alpha}(O_u(x))Q_{\alpha}(O_d(0))\rangle$. The OPE
$Q^{\alpha}(O_u(x))\times Q_{\alpha}(O_d(0))$ can be computed as follows.
One picks out the $\theta_{1}\theta_{2}$ term in the expansion of
\ChiralChiralThreePF,  as this corresponds to the three-point function
between $Q^\alpha\CO_u$, $Q_\alpha \CO_d$, and a third operator. From this,
one can extract the terms in the OPE $Q^{\alpha}(O_u(x))\times
Q_{\alpha}(O_d(0))$. The result is
\eqna{
Q^{\alpha}(O_{u}(x))Q_{\alpha}(O_{d}(0))&=
c_{\mu;1}Q^{2}O_{1,0}(0)
+ c_{\mu;2} Q^{\alpha}O_{2,1\,\alpha}(0)\\
&\quad+\sum_i(c_{\mu;3}^iO_{3,0;i}(0)
+c_{\mu;4}^i[Q^2\bar{Q}^2O_{3,0;i}]_p(0)
+c_{\mu;5}^i[Q\sigma_{\mu}\bar{Q}O_{3,1;i}^\mu]_p(0))
+\cdots,}[muOPE]
where $i$ is a counting index, with
\eqn{\begin{gathered}
c_{\mu;1}=\frac{\lambda_{\CO_u\CO_d\CO_{1,0}}}
{C_{\CO_{1,0}}}\frac{\Delta_{\CO_u}\Delta_{\CO_d}}{(\Delta_{\CO_u} +
\Delta_{\CO_d})(\Delta_{\CO_u}+\Delta_{\CO_d}-1)},\qquad
c_{\mu;2}=\frac{\lambda_{\CO_u\CO_d\CO_{2,1}}}{C_{\CO_{2,1}}}
\frac{\Delta_{\CO_u}-\Delta_{\CO_d}}{\Delta_{\CO_u}+\Delta_{\CO_d}-2},\\
c_{\mu;3}^i={\check{c}}_{\mu;3}^ix^{\Delta_{\CO_{3,0;i}}-\Delta_{\CO_u}
-\Delta_{\CO_d}-1},\qquad\!\!
c_{\mu;4}^i={\check{c}}_{\mu;4}^ix^{\Delta_{\CO_{3,0;i}}-\Delta_{\CO_u}
-\Delta_{\CO_d}+1},\qquad\!\!
c_{\mu;5}^i={\check{c}}_{\mu;5}^ix^{\Delta_{\CO_{3,1;i}^\mu}\!
-\Delta_{\CO_u}-\Delta_{\CO_d}},
\end{gathered}}[]
where
\eqna{{\check{c}}_{\mu;3}^i&=4\frac{\lambda_{\CO_u\CO_d\CO_{3,0;i}}}
{C_{\CO_{3,0;i}}},\\
{\check{c}}_{\mu;4}^i&=\frac{1}{2^4}\frac{\lambda_{\CO_u\CO_d\CO_{3,0;i}}}
{C_{\CO_{3,0;i}}}\frac{(\Delta_{\CO_u}-\Delta_{\CO_d}
-\Delta_{\CO_{3,0;i}}-1)(\Delta_{\CO_u}-\Delta_{\CO_d}
+\Delta_{\CO_{3,0;i}}+1)}{\Delta_{\CO_{3,0;i}}
(\Delta_{\CO_{3,0;i}}+1)(\Delta_{\CO_u}+\Delta_{\CO_d}
-\Delta_{\CO_{3,0;i}}-1)(\Delta_{\CO_u}+\Delta_{\CO_d}
-\Delta_{\CO_{3,0;i}}-3)},\\
{\check{c}}_{\mu;5}^i&=\frac{i}{2}\frac{\lambda_{\CO_u\CO_d
\CO^{\mu}_{3,1;i}}}{C_{\CO_{3,1;i}^\mu}}\frac{\Delta_{\CO_u}-
\Delta_{\CO_d}}{(\Delta_{\CO^{\mu}_{3,1;i}}-2)(\Delta_{\CO^{\mu}_{3,1;i}}
-\Delta_{\CO_u}-\Delta_{\CO_d})}.}[]
For a flavor of what is involved in the derivation of the $c_{\mu}^i$'s see
Appendix~\ref{technicalThreeP}. Note that the three-point function
coefficients $\lambda_{\CO_u\CO_d\CO_{2,1}}$ and $\lambda_{\CO_u\CO_d
\CO^{\mu}_{3,1;i}}$ are antisymmetric under $\CO_u \leftrightarrow \CO_d$.

Some comments are in order. We see that all three solutions in section
\ref{chiralchiral} contribute to the OPE. From \muOPE we see that a
supersymmetric contribution to $\mu$ can only arise from $O_{3,0}$, since
all other contributions arise as $Q(\cdot)$.  However, $O_{3,0}$ can also
get contributions from supersymmetry-breaking. If the theory contains
spurions, then operators constructed with them can get a vev. Such
operators correspond to $O_{3,0}$ in \muOPE. In a given model one has to
identify candidate operators that can appear in the right-hand side of
\muOPE and obtain an expectation value consistent with Poincar\'e and gauge
invariance. Those are generally not unique operators, so \muOPE contains a
sum over all appropriate operators in the right-hand side. By substituting
the above result in \HiggsParams and using dispersion relations, it is
possible to compute $\mu$. We will do this explicitly for a model in
section \ref{toy}.

Before we proceed, though, let us examine the first two terms of \muOPE
more carefully. Those terms have no $x$-dependence, which is a reflection
of the fact that the scaling dimension of the operators $Q^2 O_{1,0}$ and
$Q^\alpha O_{2,1\,\alpha}$ is determined by the scaling dimensions of the
operators in the left-hand side of \muOPE. Now, from \HiggsParams we see
that we have to Fourier-transform \muOPE and take the zero-momentum limit.
If we use the OPE to express the operator product in the two-point
function, though, the zero-momentum limit becomes problematic, for the
Wilson coefficient cannot be evaluated directly in that limit. However, we
can use the dispersion relation \integrate to argue that the terms in the
first line of \muOPE actually \emph{do not} contribute to $\mu$. Indeed, in
order to regulate the Fourier transform let us use
\eqn{i\int \dfour x\,e^{-ip\cdot x}\frac{1}{(x^2)^\epsilon}=
\pi^2\frac{\Gamma(2-\epsilon)}{2^{2\epsilon-4}\Gamma(\epsilon)}
\frac{1}{(p^2)^{2-\epsilon}},}[FT]
and take the limit $\epsilon\to 0$ at the end of the computation. Expanding
the right-hand side of \FT we see that all terms involving $\ln(-s)$ are
multiplied with at least one power of $\epsilon$, and thus applying
\integrate gives a vanishing result as $\epsilon\to 0$. Thus, the first two
terms in \muOPE do not give rise to contributions to $\mu$.

It is also possible to understand this result intuitively. The fact that
the Wilson coefficients of the operators $Q^2O_{1,0}$ and
$Q^{\alpha}O_{2,1\,\alpha}$ have no $x$-dependence is inconsistent with the
presence of a threshold at the scale $M$. Indeed, if there is no
$x$-dependence one can take $x$ to be very large (such as $x \gg
\frac{1}{M}$), and still expect a non-vanishing contribution to $\mu$.
However, from general considerations one expects that the presence of a
threshold at a scale $M$ implies that a general two-point function should
factorize, i.e.\ $\langle\CO_1(x) \CO_2(0) \rangle \sim
\langle\CO_1(x)\rangle \langle\CO_2(0)\rangle$ for $x\gg\frac{1}{M}$.  In
the example above, the relevant operators are $Q^{\alpha}O_u$ and
$Q_{\alpha}O_d$, which are assumed to not get vevs, since that would break
the visible-sector gauge group. Thus, the two-point function must die away
for $x\gg\frac{1}{M}$.  This implies that the only consistent possibility
is that these terms do not contribute, as can be verified from the
computation with \FT above.

Consequently, we can use \HiggsParams to write
\eqn{\mu=\lambda_u\lambda_d\sum_i ({\hat{c}}_{\mu;3}^i\langle O_{3,0;i}
\rangle+{\hat{c}}_{\mu;4}^i\langle Q^2\bar{Q}^2O_{3,0;i}\rangle
+{\hat{c}}_{\mu;5}^i\langle Q\sigma_\mu\bar{Q}O_{3,1;i}^\mu\rangle),
}[muterm]
with
\eqna{{\hat{c}}_{\mu;3}^i&=\frac{i}{8}{\check{c}}_{\mu;3}^i\left.\int\dfour
x\,e^{-ip\cdot x}x^{\Delta_{\CO_{3,0;i}}-\Delta_{\CO_u}-\Delta_{\CO_d}-1}\right|_{p\to0},\\
{\hat{c}}_{\mu;4}^i&=\frac{i}{8}{\check{c}}_{\mu;4}^i
\left.\int\dfour x\,e^{-ip\cdot x}x^{\Delta_{\CO_{3,0;i}}-\Delta_{\CO_u}-\Delta_{\CO_d}+1}\right|_{p\to0},\\
{\hat{c}}_{\mu;5}^i&=\frac{i}{8}{\check{c}}_{\mu;5}^i
\left.\int\dfour x\,e^{-ip\cdot x}x^{\Delta_{\CO_{3,1;i}^\mu}\!
-\Delta_{\CO_u}-\Delta_{\CO_d}}\right|_{p\to0}.}[muchat]
As stated in the introduction, the two-point functions in \HiggsParams and
\HiggsHard are assumed to have a branch cut, starting from a threshold
$s_0=4M^2$ in momentum space, and no other singularities.  Each Wilson
coefficient in \muOPE has a branch cut in the UV determined by
superconformal symmetry.  Its branch point in general differs from $s_0$.
After resumming the OPE, branch cuts from Wilson coefficients should
combine into that of the two-point function.  In cases where such complete
resummation is not practical, the first few terms in the OPE may still
provide a reasonable estimate, if their branch cuts are assumed to start
from the threshold.  This further approximation provides an IR regulator
for evaluating \muchat.  Fourier transforming a Wilson coefficient with
\FT, we may obtain its value at $p^2\rightarrow0$ by integrating around
its branch cut that starts from $s_0$. The result is
\eqn{\lim_{p^{2}\rightarrow0}i\int\dfour x\,e^{-ip\cdot x}x^{\alpha}
\rightarrow \frac{\alpha+2}{\alpha+4}
\Gamma^2\!\left(1+\frac{\alpha}{2}\right)\sin^{2}
\left(\frac{\alpha\pi}{2}\right)\frac{1}{M^{\alpha+4}}.}[pToZero]
Applying this result to \muchat we get
\eqna{
\hat{c}_{\mu;3}^{i}&=\frac{\lambda_{\mathcal{O}_{u}\mathcal{O}_{d}
\mathcal{O}_{3,0;i}}}{C_{\mathcal{O}_{3,0;i}}}\frac{\eta_{0}+1}{\eta_{0}+3}
\Gamma^2\!\left(\frac{\eta_{0}+1}{2}\right)
\sin^{2}\left(\frac{\eta_{0}\pi}{2}\right)\frac{1}{M^{\eta_{0}+3}},\\
\hat{c}_{\mu;4}^{i}&=\frac{1}{2^{9}}\frac{\lambda_{\mathcal{O}_{u}
\mathcal{O}_{d}\mathcal{O}_{3,0;i}}}{C_{\mathcal{O}_{3,0;i}}}
\frac{(\Delta_{\mathcal{O}_{u}}-\Delta_{\mathcal{O}_{d}}-
\Delta_{\mathcal{O}_{3,0;i}}-1)(\Delta_{\mathcal{O}_{u}}-
\Delta_{\mathcal{O}_{d}}+\Delta_{\mathcal{O}_{3,0;i}}+1)}
{\Delta_{\mathcal{O}_{3,0;i}}(\Delta_{\mathcal{O}_{3,0;i}}+1)}\times\\
&\hspace{7cm}\frac{\eta_{0}+1}{\eta_{0}+5}\Gamma^2\!\left(\frac{\eta_{0}+1}
{2}\right)\cos^{2}\left(\frac{\eta_{0}\pi}{2}\right)
\frac{1}{M^{\eta_{0}+5}},\\
\hat{c}_{\mu;5}^{i}&=-\frac{i}{2^{4}}\frac{\lambda_{\mathcal{O}_{u}
\mathcal{O}_{d}\mathcal{O}_{3,1;i}}}{C_{\mathcal{O}_{3,1;i}}}
\frac{\Delta_{\mathcal{O}_{u}}-\Delta_{\mathcal{O}_{d}}}
{\Delta_{\mathcal{O}_{3,1;i}^{\mu}}-2}
\frac{\eta_{1}+2}{\eta_{1}(\eta_{1}+4)}
\Gamma^2\!\left(\frac{\eta_{1}}{2}+1\right)
\sin^{2}\left(\frac{\eta_{1}\pi}{2}\right)\frac{1}{M^{\eta_{1}+4}},
}[muchatresult]
where
\eqn{\eta_{0}\equiv\Delta_{\mathcal{O}_{3,0;i}}-\Delta_{\mathcal{O}_{u}}
-\Delta_{\mathcal{O}_{d}},\qquad
\eta_{1}\equiv\Delta_{\mathcal{O}_{3,1;i}^{\mu}}-\Delta_{\mathcal{O}_{u}}
-\Delta_{\mathcal{O}_{d}}.}[etas]

Equivalently, the calculation of the necessary OPE coefficients can be done
directly in momentum space. We will see an explicit example of the latter
approach in section \ref{toy}.

\subsubsection{\texorpdfstring{$B_{\mu}$}{Bmu}}
The parameter $B_{\mu}$ is determined by $\langle
Q^2(O_u(x))Q^2(O_d(0))\rangle$.  Similar to the previous case, one has to
now expand the three-point function \ChiralChiralThreePF to order
$\theta_{1}^{2} \theta_{2}^{2}$, which corresponds to the three-point
function between $\tfrac14Q^2\CO_u$, $\tfrac14Q^2\CO_d$, and a third
operator. From this, one can extract the terms in the OPE
$Q^2\CO_u(x)\times Q^2\CO_d(0)$.

It turns out that only \ref{SolThree} with spin $\ell=0$ contributes to the
$B_{\mu}$ term. This can be understood as follows. For a potential
\ref{SolOne} contribution, every $\theta_{1,2}$ has to come with a
$\bar{\theta}_{3}$.  But since $\bar{\theta}_{3}^{\hspace{0.5pt}4}$
vanishes, \ref{SolOne} does not contribute.  For a potential \ref{SolTwo}
contribution with $\ell=1$, one $\theta_{1,2}$ could come from
$\bar{\Theta}_{3}$, but there are at least three $\theta_{1,2}$'s which
need to be paired with $\bar{\theta}_{3}$. Again, since
$\bar{\theta}_{3}^{\hspace{0.5pt}3}$ vanishes, \ref{SolTwo} does not
contribute as well.  Finally, for \ref{SolThree} with $\ell=1$ the lowest
scalar component field arises at order
$\theta_{3}\sigma_{\mu}\bar{\theta}_{3}$.  So one needs at least order
$\theta_{1}^{2}\theta_{2}^{2} \theta_{3}
\bar{\theta}_{3}^{\hspace{0.5pt}3}$ because two $\theta$'s  could come from
$\bar{\Theta}^{2}$.  Again, since $\bar{\theta}_{3}^{\hspace{0.5pt}3}$
vanishes, \ref{SolThree} with $\ell=1$ does not contribute. Of course these
results can be understood from the relation between $\mu$ and $B_\mu$ we
mentioned after \HiggsParams.

The contribution from \ref{SolThree} with $\ell=0$ is
\eqn{Q^{2}(O_u(x))Q^{2}(O_d(0))=\sum_ic_{B_\mu}^iQ^2O_{3,0;i}(0),\qquad
c_{B_\mu}^i=-4\frac{\lambda_{\CO_u\CO_d\CO_{3,0;i}}}{C_{O_{3,0;i}}}
x^{\Delta_{\CO_{3,0;i}}-\Delta_{\CO_u}-\Delta_{\CO_d}-1}=-c_{\mu,3}^i,
}[BmuOPE]
which allows us to compute $B_\mu$ using \HiggsParams, with the result
\eqn{B_\mu=\lambda_u\lambda_d\sum_i \hat{c}_{B_\mu}^i\langle
Q^2O_{3,0;i}\rangle,}[Bmuterm]
where
\eqn{\hat{c}_{B_\mu}^i=-\tfrac14\hat{c}_{\mu;3}^i.}[BmumuWilson]
To summarize, $B_\mu$ receives contributions only from $Q^2O_{3,0}$. This
is to be contrasted with $\mu$, which receives contributions from
$O_{3,0}$, $Q^2\bar{Q}^2O_{3,0}$, and $Q\sigma_{\mu}\bar{Q}O_{3,1}^\mu$.
The implications of this result for solutions to the $\mu/B_\mu$ problem in
models of gauge mediation will be discussed more in section \ref{discuss}.

\subsubsection{\texorpdfstring{$\delta A_{u,d}$}{delta A} and
\texorpdfstring{$\delta m_{H_{u,d}}^2$}{delta mH-squared}}
We now study correlation functions which determine the chiral-antichiral
OPE. From \HiggsParams, we see that $\delta A_{u,d}$ and $\delta
m^2_{H_{u,d}}$ are determined by $\langle Q^{2}(O_{u,d}(x)
O_{u,d}^{\dagger}(0))\rangle $ and $\langle Q^{2}\bar{Q}^{2}(O_{u,d}(x)
O_{u,d}^{\dagger}(0))\rangle$ respectively. Thus, we carry out the OPE
$O_{u,d}(x) \times O_{u,d}^{\dagger}(0)$, and then act with $Q^2$ and
$Q^2\bar{Q}^2$ to obtain $\delta A_{u,d}$ and $\delta m_{H_{u,d}}^2$
respectively. For this computation we can set
$\theta_{1,2}=\bar{\theta}_{1,2}=0$,  in order to focus on the three-point
function of $O_{u,d}$, $O_{u,d}^\dagger$ and a third operator. From this,
one can extract terms in the OPE $O_{u,d}(x)\times O_{u,d}^{\dagger}(0)$ as
before.

Now, it can be checked that the OPE $O_{u,d}(x)\times O_{u,d}^\dagger(0)$
receives contributions from various components of a general superfield
$\CO_0$ (the subscript stands for the spin $\ell=0$).  The possibilities
are the scalar component $O_0$, the $\theta\sigma^{\mu}\bar{\theta}$
component $V_{\CO_0}^{\mu}$, and the $\theta^2\bar{\theta}^2$ component
$D_{\CO_0}$.  However, we are ultimately interested in $\delta A_{u,d}$ and
$\delta m^2_{H_{u,d}}$, which are obtained by acting on the above with
$Q^2$ and $Q^2\bar{Q}^2$ respectively and taking the vev. The only operator
of interest after this is $O_0$.

The superfield three-point function for the $\ell=0$ case is
\eqn{\langle \CO_{u,d}(z_{1+})\bar{\CO}_{u,d}(z_{2-}){\bar{\CO}}_0(z_{3})
\rangle=\lambda_{\CO_{u,d}\bar{\CO}_{u,d}\CO_{0}}
\frac{\osbx{\bar{2}1}{\Delta_{\CO_0}
-2\Delta_{\CO_{u,d}}}}{\osbx{\bar{2}3}{\Delta_{\CO_0}}
\osbx{\bar{3}1}{\Delta_{\CO_0}}},}[]
from which one can extract the OPE
\eqn{O_{u,d}(x)O_{u,d}^{\dagger}(0)=c_{u,d}^i O_{0;i}^{u,d}(0)+\cdots,
\qquad c_{u,d}^i={\check c}_{u,d}^i
x^{\Delta_{\CO_{0;i}^{\smash{u,d}}}-2\Delta_{\CO_{u,d}}},\qquad
{\check c}_{u,d}^i=\frac{\lambda_{\CO_{u,d}\bar{\CO}_{u,d}
\CO_{0;i}^{u,d}}}{C_{\CO_{0;i}^{u,d}}}.}[amh2OPE]
With this result we can compute $\delta A_{u,d}$ and $\delta m_{H_{u,d}}^2$
using expressions \HiggsParams. We obtain
\eqn{\delta A_{u,d}=|\lambda_{u,d}|^2\sum_i {\hat{c}}_{A_{u,d}}^i
\langle Q^2 O_{0;i}^{u,d}\rangle,\qquad
{\hat c}_{A_{u,d}}^i=-\frac{i}{8}{\check c}_{u,d}^i\left.\int\dfour x\,
e^{-ip\cdot x} x^{\Delta_{\CO_{0;i}^{\smash{u,d}}}-2\Delta_{\CO_{u,d}}}
\right|_{p\to0}}[Aterm]
If we evaluate ${\hat c}_{A_{u,d}}^i$ with \pToZero, we get
\eqn{\hat{c}_{A_{u,d}}^{i}=-\frac{1}{2^{3}}\frac{\gamma_{u,d}+2}
{\gamma_{u,d}+4}\Gamma^2\!\left(1+\frac{\gamma_{u,d}}{2}\right)
\sin^{2}\left(\frac{\gamma_{u,d}\pi}{2}\right)
\frac{\lambda_{\mathcal{O}_{u,d}\mathcal{O}_{u,d}^{\dagger}
\mathcal{O}_{0;i}^{u,d}}}{C_{\mathcal{O}_{0;i}^{u,d}}}
\frac{1}{M^{\gamma_{u,d}+4}},}[]
where
\eqn{\gamma_{u,d}=\Delta_{\CO_{0;i}^{u,d}}-2\Delta_{\CO_{u,d}}.}
The $\delta m_{H_{u,d}}^2$ term is given by
\eqn{\delta m_{H_{u,d}}^2=|\lambda_{u,d}|^2\sum_i {\hat
c}_{m_{H_{u,d}}^2}^i\langle Q^2\bar{Q}^2 O_{0,i}^{u,d}\rangle,\qquad
{\hat c}_{m_{H_{u,d}}^2}^i\!\!=
-\frac{i}{2^5}{\check c}_{u,d}^i\left.\int\dfour x\,e^{-ip\cdot x}
x^{\Delta_{\CO_{0;i}^{\smash{u,d}}}-2\Delta_{\CO_{u,d}}}\right|_{p\to0}.
}[mHterm]
Note that, unlike the case of $\mu$ and $B_{\mu}$, the same term in the OPE
contributes to both $\delta A_{u,d}$ and $\delta m_{H_{u,d}}^2$. Analogous
to \BmumuWilson we here have the relation
\eqn{{\hat c}_{m_{H_{u,d}}^2}^i\!\!=\tfrac14 {\hat
c}_{A_{u,d}}^i.}[mHAWilson]
This will have implications for the $A/m_{H}^2$
problem~\cite{Craig:2013wga}, and will also be discussed in section
\ref{discuss}.

\subsubsection{\texorpdfstring{$a'_{u,d}$}{a'} (``wrong Higgs''
trilinears)}
It is also possible to compute the couplings in
$V^{(\text{other})}_{\text{Higgs}}$ (see \HiggsHard) using the same
methods. As mentioned earlier, and explained in~\cite{Komargodski:2008ax},
$\delta \mathcal{Z}_{u,d}$ arises from supersymmetric wave-function
renormalization and does not affect any physical observables at leading
order in $\lambda_{u,d}$. Next, we study the couplings $a'_{u,d}$.  As seen
from \LHiggs, \HiggsParams and \HiggsHard, $a'_{u,d}$ arises from effects
which cannot be captured by the $\mu$ parameter alone.  They can be thought
of as providing ``wrong Higgs'' trilinear parameters after one replaces
$F_{\CH_u}$ and $F_{\CH_d}$ by their equation of motion (see
\LHiggs).\footnote{The $a'_{u,d}$ terms are called ``maybe soft''
in~\cite{Martin:1997ns}, but if there are no SM gauge singlets in the
theory (as in the MSSM), they are soft and do not introduce quadratic
divergences (see discussion in~\cite{Martin:1997ns}).}

From \HiggsHard, $a'_{u,d}$ is determined by $\langle
Q^\alpha(O_{u,d}(x)\,Q_\alpha(O_{d,u}(0)))\rangle$. Since it can
be written as $\langle Q(\cdot)\rangle$, it can only get contributions
starting at order $F/M$.\footnote{Here we assume that $F/M^2$ is smaller
than unity, as mentioned earlier.} In order to compute $a'_{u,d}$, we first
find the OPE $O_{u,d}(x)\times Q_\alpha(O_{d,u}(0))$ and then act with
$Q^{\alpha}$. Using the same procedure as for the other parameters, we find
\eqna{Q^{\alpha}(O_{u,d}(x)Q_{\alpha}O_{d,u}(0))&=
c_{a_{u,d;1}^\prime}Q^{2}O_{1,0}(0)
+c_{a_{u,d;2}^\prime}Q^{\alpha}O_{2,1\,\alpha}(0)\\
&\quad+\sum_i(c_{a_{u,d;3}^\prime}^i[Q^{2}\bar{Q}^{2}O_{3,0;i}]_p(0)
+c_{a_{u,d;4}^\prime}^i[Q\sigma_{\mu}\bar{Q}O_{3,1;i}^\mu]_p(0))+
\cdots.}[aprimeOPE]
with\foot{We thank J.-F.\ Fortin for bringing to our attention a typo in
$c_{a_{u,d;2}^\prime}$ and a mistake in ${\check{c}}_{a_{u,d;3}^\prime}^i$
in previous versions of this paper, before publication of his
work~\cite{Fortin:2017rzx}.}
\eqn{\begin{gathered}
c_{a_{u,d;1}^\prime}=\frac{\Delta_{\CO_{d,u}}}{\Delta_{\CO_u}
+\Delta_{\CO_d}}\frac{\lambda_{\CO_{u,d}\CO_{d,u}\CO_{1,0}}}
{C_{\CO_{1,0}}},\qquad
c_{a_{u,d;2}^\prime}=-\frac{\lambda_{\CO_{u,d}\CO_{d,u}\CO_{2,1}}}
{C_{\CO_{2,1}}},\\
c_{a_{u,d;3}^\prime}^i={\check{c}}_{a_{u,d;3}^\prime}^i
x^{\Delta_{\CO_{3,0;i}}-\Delta_{\CO_u}-\Delta_{\CO_d}+1},\qquad
c_{a_{u,d;4}^\prime}^i={\check{c}}_{a_{u,d;4}^\prime}^i
x^{\Delta_{\CO_{3,1;i}^\mu}\!-\Delta_{\CO_u}-\Delta_{\CO_d}},
\end{gathered}}[]
where
\eqna{{\check{c}}_{a_{u,d;3}^\prime}^i&=\frac18\frac{\lambda_{\CO_{u,d}
\CO_{d,u}\CO_{3,0;i}}}{C_{\CO_{3,0;i}}}\\
&\hspace{-0.5cm}\times\frac{(\Delta_{\CO_{3,0;i}}-1)
(\Delta_{\CO_{u,d}}-\Delta_{\CO_{d,u}}-\Delta_{\CO_{3,0;i}}-1)}
{(\Delta_{\CO_{3,0;i}}+1)(\Delta_{\CO_u}+\Delta_{\CO_d}
-\Delta_{\CO_{3,0;i}}-1)(\Delta_{\CO_u}+\Delta_{\CO_d}
-\Delta_{\CO_{3,0;i}}-3)(\Delta_{\CO_u}+\Delta_{\CO_d}
+\Delta_{\CO_{3,0;i}}-3)},\\
{\check{c}}_{a_{u,d;4}^\prime}^i&=\frac{i}{2}
\frac{\lambda_{\CO_{u,d}\CO_{d,u}\CO_{3,1;i}^\mu}}
{C_{\CO_{3,1;i}^\mu}}\frac{1}{
\Delta_{\CO_u}+\Delta_{\CO_d}-\Delta_{\CO_{3,1;i}^\mu}}.}[]

Just like in the case of $\mu$, the first two terms in the right-hand side
of \aprimeOPE do not end up contributing to $a'_{u,d}$.  Therefore, we can
write
\eqn{a'_{u,d}=\sum_i({\hat{c}}_{a_{u,d;3}^\prime}^i
\langle Q^2\bar{Q}^2O_{3,0;i}\rangle
+{\hat{c}}_{a_{u,d;4}^\prime}^i
\langle Q\sigma_\mu\bar{Q}O_{3,1;i}^\mu\rangle),}[aprimeterm]
with
\eqna{
{\hat{c}}_{a_{u,d;3}^\prime}^i&=\frac{i}{8}{\check{c}}_{a_{u,d;3}^\prime}^i
\left.\int\dfour x\,e^{-ip\cdot x}x^{\Delta_{\CO_{3,0;i}}-\Delta_{\CO_u}-\Delta_{\CO_d}+1}\right|_{p\to0},\\
{\hat{c}}_{a_{u,d;4}^\prime}^i&=\frac{i}{8}{\check{c}}_{a_{u,d;4}^\prime}^i
\left.\int\dfour x\,e^{-ip\cdot x}x^{\Delta_{\CO_{3,1;i}^\mu}\!
-\Delta_{\CO_u}-\Delta_{\CO_d}}\right|_{p\to0}.}[]
Using \pToZero we get
\eqna{\hat{c}_{a_{u,d;3}^{\prime}}^{i}&=\frac{1}{2^{8}}
\frac{\lambda_{\mathcal{O}_{u,d}\mathcal{O}_{d,u}\mathcal{O}_{3,0;i}}}
{C_{\mathcal{O}_{3,0;i}}}\frac{\Delta_{\mathcal{O}_{u,d}}
-\Delta_{\mathcal{O}_{d,u}}-\Delta_{\mathcal{O}_{3,0;i}}-1}
{\Delta_{\mathcal{O}_{3,0;i}}+1}\times\\
&\hspace{6cm}\frac{\eta_{0}+1}{\eta_{0}+5}
\Gamma^2\!\left(\frac{\eta_{0}+1}{2}\right)
\cos^{2}\left(\frac{\eta_{0}\pi}{2}\right)
\frac{1}{M^{\eta_{0}+5}},\\
\hat{c}_{a_{u,d;4}^{\prime}}^{i}&=-\frac{i}{2^{4}}
\frac{\lambda_{\mathcal{O}_{u,d}\mathcal{O}_{d,u}\mathcal{O}_{3,1;i}}}
{C_{\mathcal{O}_{3,1;i}}}\frac{\eta_{1}+2}{\eta_{1}(\eta_{1}+4)}
\Gamma^2\!\left(\frac{\eta_{1}}{2}+1\right)
\sin^{2}\left(\frac{\eta_{1}\pi}{2}\right)\frac{1}{M^{\eta_{1}+4}},
}[chataprime]
with $\eta_{0,1}$ given in \etas.

By comparing \muterm and \aprimeterm, we see that $Q^2\bar{Q}^2O_{3,0}$ and
$Q\sigma_{\mu}\bar{Q} O_{3,1}^\mu$ contribute to both $\mu$ and $a'_{u,d}$,
with comparable Wilson coefficients. This implies that unless $O_{3,0}$ is
the dominant operator contributing to $\mu$, $a'_{u,d}$ may be generated at
the same order as $\mu$ in general. In simple (spurion-based) models of
supersymmetry breaking, however, it can be shown that $O_{3,0}$ gives the
dominant contribution to $\mu$. Hence, in such cases, $a'_{u,d}$ is
suppressed compared to $\mu$.

Note that although $a'_{u,d}$ may be generated at the same order as $\mu$
in general, it can still only be generated at order $F^2$. This is because
both operators $Q^{2}\bar{Q}^{2}O_{3,0}$ and
$Q\sigma_{\mu}\bar{Q}O_{3,1}^\mu$, which contribute to $a'_{u,d}$ in
general, are generated at order $F^2$. This can be easily seen for the
former. For the latter, this is true because $\langle
Q\sigma_{\mu}\bar{Q}O_{3,1}^\mu\rangle$ is of the form
$\langle\{Q,[\bar{Q},(.)]\}\rangle$, and all such correlation functions of
this form must start at order $F^2$~\cite{Komargodski:2008ax}. Thus, the
wrong-Higgs trilinears will generically be suppressed to other soft
parameters which start at order $F$. We will make more comments about this
result in section \ref{discuss}.

\subsubsection{\texorpdfstring{$\gamma$}{gamma}}
Finally, we discuss the remaining parameter in
$V^{(\text{other})}_{\text{Higgs}}$, namely $\gamma$, which gives rise to a
dimensionless coupling between four MSSM sfermions for example (in addition
to other terms), after using the equation of motion (see \LHiggs). Now,
$\gamma$ is determined by $\langle O_u(x) O_d(0)\rangle$ from \HiggsHard.
It can be shown that this dimensionless parameter must be suppressed at
least by order $F/M^2$ in phenomenologically viable models.

Indeed, suppose $\langle \CO_u(x)\CO_d(0)\rangle$ is nonvanishing in the
supersymmetric limit. Then, since it is a correlation function of scalar
chiral primary operators, it does not depend on the separation $|x|$ of the
two operators~\cite{Amati:1988ft}. This implies that one can take $|x|$ to
be very large, and apply the cluster-decomposition principle. Thus,
$\lim_{|x|\rightarrow \infty}\,\langle O_u(x)O_d(0)\rangle = \langle
O_u\rangle \langle O_d\rangle \neq 0$. However, since $O_u$ and $O_d$ are
charged under the SM gauge symmetry (see \W), this explicitly breaks the SM
gauge symmetry. Hence, $\langle \CO_u(x)\CO_d(0)\rangle$ can only receive
contributions from supersymmetry breaking, and must be at least of order
$F/M^2$.

\newsec{A weakly-coupled example}[toy]
In this section we use OPE methods to compute the Higgs parameters for a
simple model that was considered in~\cite{Dvali:1996cu}. This model
illustrates the $\mu/B_\mu$ problem in gauge-mediated supersymmetry
breaking. The model contains messengers $\Phi_{1,2}$ and
$\widetilde{\Phi}_{1,2}$, with $\CO_u=\Phi_1\Phi_2$ and
$\CO_d=\widetilde{\Phi}_1\widetilde{\Phi}_2$.  Here $\Phi_1$ is a
$\textbf{5}$, $\widetilde{\Phi}_1$ a $\bar{\textbf{5}}$, and $\Phi_2$ and
$\widetilde{\Phi}_2$ singlets of $SU(5)$. The superpotential that couples
the messenger to the Higgs sector is
\eqn{W=\lambda_u\CH_u\Phi_1\Phi_2+\lambda_d\CH_d \widetilde{\Phi}_1
\widetilde{\Phi}_2=\lambda_u\epsilon^{ij}(H_u)_i(\Phi_1)_j\Phi_2 +
\lambda_d\epsilon_{ij}(H_d)^i(\widetilde{\Phi}_1)^j\widetilde{\Phi}_2,}[]
where $i,j$ are $SU(2)$ indices, while the messenger sector is coupled to
the hidden sector via
\eqn{W=\lambda X(\Phi_1\widetilde{\Phi}_1+\Phi_2\widetilde{\Phi}_2),}[]
where $X$ is a spurion that gets a vev in both the first and the last
component, $\langle X\rangle=X+\theta^2 F$. The computations below are done
at one loop and to leading order in $F/X^2$.

To start, let us compute the leading contribution to $\mu$. From
\HiggsParams we see that we have to consider the OPE of the fermionic
components of the composite operators $\Phi_1\Phi_2$ and
$\widetilde{\Phi}_1 \widetilde{\Phi}_2$.  We use the results of section
\ref{computations} to compute the right-hand side of the OPE, and
concentrate on the operator of the lowest scaling dimension that has a
nonzero supersymmetry-breaking vev consistent with Poincar\'e and gauge
invariance:
\eqn{i\int\dfour x\,e^{-ip\cdot x}\,Q^\alpha(\Phi_1\Phi_2(x))Q_\alpha
(\widetilde{\Phi}_1 \widetilde{\Phi}_2(0))= \tilde{c}_{X^\dagger
F^{\dagger}}(s)X^\dagger F^\dagger(0)+\cdots,\qquad s=-p^2.}[]
Note that $Q^\alpha(\Phi_1\Phi_2)=-i\sqrt{2}(\Phi_1\psi_{\Phi_2}^\alpha
+\Phi_2\psi_{\Phi_1}^\alpha)$ and similarly for $\widetilde{\Phi}_{1,2}$.
The operator $X^\dagger F^\dagger$, which is an example of an $\CO_{3,0}$
in \muOPE, turns out to be the leading operator with the correct
$R$-charge. Of course, there are higher dimension operators of the form
$X^{\dagger}F^{\dagger} (X^{\dagger}X)^n (F^{\dagger}F)^m$, with $n,m\geq0$
but not both zero, which also contribute to the OPE. However, as advocated
in~\cite{Fortin:2012tp} we expect a good approximation to the full answer
with this truncation. The Wilson coefficient at one loop is given by
(messengers run in the loop)\foot{The factor of $-2^3$ comes about as
follows: a 2 because we have two equal diagrams due of the presence of both
$\Phi_{1,2}$ and $\widetilde{\Phi}_{1,2}$, a $-2=(-i\sqrt{2})^2$ from the
insertions of the operators at the crosses, and a 2 from the fact that our
messengers $\Phi_1$ and $\widetilde{\Phi}_1$ are in the $\mathbf{5}$ and
$\bar{\mathbf{5}}$ of $SU(5)$ respectively, with the branching rule
$\mathbf{5}=(\mathbf{1}, \mathbf{2})_{-\frac12} +(\mathbf{1},
\mathbf{3})_{\frac13}$ into the standard model $SU(3)\times SU(2)\times
U(1)$, so that doublets run in the loop.}
\eqn{-i\tilde{c}_{X^\dagger F^\dagger}(s)=-2^3\times\hspace{-4pt}
\begin{tikzpicture}[>=latex',baseline=(vert_cent.base)]
  \node (vert_cent) {$\phantom{\cdot}$};
  \tikzstyle{arr}=[decoration={markings,mark=at position 1 with
      {\arrow[scale=1.5]{>}}}, postaction={decorate}];
  \draw[arr] (0,0)--(0.6,0) node[midway,above] {$p$};
  \draw[arr] (3.4,0)--(4,0) node[midway,above] {$p$};
  \draw (2,0) ++(0:1cm) arc (0:180:1cm);
  \draw[dashed,dash phase=-1pt] (2,0) ++(0:-1cm) arc (180:360:1cm);
  \draw[arr] (2,0) ++(120:1.2cm) arc (120:150:1.2cm)--++(-122:1pt)
    node[midway,above=5pt] {$\,k$};
  \filldraw[cross,fill=white] (1,0) circle (4pt);
  \filldraw[cross,fill=white] (3,0) circle (4pt);
  \draw[dashed,double,dash phase=4pt] (2,-1)--(2,-1.8) node[below]
    {$F^\dagger$};
  \draw[dashed,dash phase=4pt] (2,1)--(2,1.8) node[above] {$X^\dagger$};
  \filldraw[black] (2,1) circle (1pt) node[below] {$\lambda^\ast$};
  \filldraw[black] (2,-1) circle (1pt) node[above] {$\lambda^\ast$};
\end{tikzpicture}\,
=\frac{i(\lambda^\ast)^2}{\pi^2}\frac{1}{Q^2}\int_0^1 dx\,x
\frac{x(1-x)+2\xi}{(x(1-x)+\xi)^2},}[muWilCoeff]
where $\xi=\mu^2/Q^2$ with $\mu$ an arbitrary normalization point necessary
in the OPE~\cite{Novikov:1984rf}, and $Q^2=p_E^2=p^2=-s$ ($s>0$ in the
physical region). Note that for convergence of the loop-integral we go to
the Euclidean region $Q^2>0$. The integral over the Feynman parameter $x$
can be performed analytically, and since we will rely on \integrate we can
expand the answer and keep only the $\ln\xi$-term. From \muWilCoeff we thus
find
\eqn{\tilde{c}_{X^\dagger F^\dagger}(s)=\frac{(\lambda^\ast)^2}{2\pi^2}
\frac{1}{s}\ln\frac{-s}{\mu^2}+\cdots,}[muWilCoeffOne]
and since $\mathrm{Im}\ln(-(s\pm i\epsilon))=\mp\pi$, $s>0$, we can use
\integrate with $s_0=4|\lambda X|^2$ to obtain
\eqn{\tilde{c}_{X^\dagger F^\dagger}(0)=-
\frac{\lambda^\ast}{4\pi^2\lambda|X|^2}.}[]
With this result it is straightforward to find the leading contribution to
$\mu$:
\eqn{\mu\approx-\frac12\frac{\lambda_u\lambda_d\lambda^\ast}
{16\pi^2\lambda}\frac{F^\dagger}{X}.}[approxmu]
This is 50\% of the result obtained in~\cite{Dvali:1996cu}. The full result
can be obtained by computing the contributions of all operators of the form
$X^{\dagger}F^{\dagger}\,(X^{\dagger}X)^n\,(F^{\dagger}F)^m$, $m,n\ge0$,
and resumming them. For simplicity, we will not do this resummation for
$\mu$ or $B_{\mu}$, but we will do part of it for $A_{u,d}$ and
$m^2_{H_{u,d}}$.

The calculation of $B_\mu$ proceeds along similar lines. As seen in
\HiggsParams we consider here the OPE $Q^2(\Phi_1\Phi_2(x))\times
Q^2(\widetilde{\Phi}_1 \widetilde{\Phi}_2(0))$. Clearly, the operator
$(X^\dagger)^2$ is the leading contribution to this OPE. Nevertheless, the
vev of this operator is supersymmetry-preserving, which means that its Wilson
coefficient must be zero, for the only contributions to $B_{\mu}$ are of
the form $Q^2(O_{3,0})$ (see \BmuOPE). Indeed, at one loop,
\eqn{-i\tilde{c}_{(X^\dagger)^2}(s)=2^5\times\hspace{-3pt}
\begin{tikzpicture}[>=latex',baseline=(vert_cent.base)]
  \node (vert_cent) {$\phantom{\cdot}$};
  \tikzstyle{arr}=[decoration={markings,mark=at position 1 with
      {\arrow[scale=1.5]{>}}}, postaction={decorate}];
  \draw[arr] (0,0)--(0.6,0) node[midway,above] {$p$};
  \draw[arr] (3.4,0)--(4,0) node[midway,above] {$p$};
  \draw (2,0) ++(0:1cm) arc (0:180:1cm);
  \draw (2,0) ++(0:-1cm) arc (180:360:1cm);
  \draw[arr] (2,0) ++(120:1.2cm) arc (120:150:1.2cm)--++(-122:1pt)
    node[midway,above=5pt] {$\,k$};
  \filldraw[cross,fill=white] (1,0) circle (4pt);
  \filldraw[cross,fill=white] (3,0) circle (4pt);
  \draw[dashed,dash phase=4pt] (2,-1)--(2,-1.8) node[below] {$X^\dagger$};
  \draw[dashed,dash phase=4pt] (2,1)--(2,1.8) node[above] {$X^\dagger$};
  \filldraw[black] (2,1) circle (1pt) node[below] {$\lambda^\ast$};
  \filldraw[black] (2,-1) circle (1pt) node[above] {$\lambda^\ast$};
\end{tikzpicture}\,
+2^6\times\hspace{-3pt}
\begin{tikzpicture}[>=latex',baseline=(vert_cent.base)]
  \node (vert_cent) {$\phantom{\cdot}$};
  \tikzstyle{arr}=[decoration={markings,mark=at position 1 with
      {\arrow[scale=1.5]{>}}}, postaction={decorate}];
  \draw[arr] (0,0)--(0.6,0) node[midway,above] {$p$};
  \draw[arr] (3.4,0)--(4,0) node[midway,above] {$p$};
  \draw[dashed] (2,0) ++(0:1cm) arc (0:90:1cm);
  \draw[dashed,double] (2,1) ++(0:0cm) arc (90:180:1cm);
  \draw[dashed] (2,0) ++(0:-1cm) arc (180:270:1cm);
  \draw[dashed,double] (3,-1) ++(0:-1cm) arc (270:360:1cm);
  \draw[arr] (2,0) ++(120:1.2cm) arc (120:150:1.2cm)--++(-122:1pt)
    node[midway,above=5pt] {$\,k$};
  \filldraw[cross,fill=white] (1,0) circle (4pt);
  \filldraw[cross,fill=white] (3,0) circle (4pt);
  \draw[dashed,dash phase=4pt] (2,-1)--(2,-1.8) node[below] {$X^\dagger$};
  \draw[dashed,dash phase=4pt] (2,1)--(2,1.8) node[above] {$X^\dagger$};
  \filldraw[black] (2,1) circle (1pt) node[below] {$\lambda^\ast$};
  \filldraw[black] (2,-1) circle (1pt) node[above] {$\lambda^\ast$};
\end{tikzpicture}\,
=0.}[BmuCancelation]
This is a consequence of supersymmetry. Similar cancellations persist
order-by-order in perturbation theory, and also for the operators
$(X^\dagger)^2(X^\dagger X)^{n}$ for any $n\geq1$.

We saw before that the leading operator of type $\CO_{3,0}$ is
$X^{\dagger}F^{\dagger}$. However, $Q^2(X^{\dagger}F^{\dagger})$ vanishes
in the zero-momentum limit. The leading operator of type $Q^2(O_{3,0})$
which contributes to $B_{\mu}$ is $Q^2(X^\dagger F^{\dagger} (X^\dagger X))
= 4(X^{\dagger})^2\,F^{\dagger} F$. The Wilson coefficient of
$(X^{\dagger})^2\,F^{\dagger} F$ at one loop is given by
\eqn{-\frac{i}{2^6}\tilde{c}_{(X^\dagger)^2F^\dagger F}(s)=\hspace{-3pt}
\begin{tikzpicture}[>=latex',baseline=(vert_cent.base)]
  \node (vert_cent) {$\phantom{\cdot}$};
  \tikzstyle{arr}=[decoration={markings,mark=at
    position 1 with {\arrow[scale=1.5]{>}}}, postaction={decorate}];
  \draw[arr] (0,0)--(0.6,0) node[midway,above] {$p$}; \draw[arr]
    (3.4,0)--(4,0) node[midway,above] {$p$};
  \draw[dashed] (2,0) ++(0:1cm) arc (0:90:1cm);
  \draw[dashed,double,dash phase=4pt] (2,1)--(2,1.8)
    node[above] {$F^\dagger$};
  \draw[dashed] (1,1) ++(0:1cm) arc (90:180:1cm);
  \draw[dashed,double] (2,0) ++(0:-1cm) arc (180:225:1cm)
    node (n1) {};
  \draw[dashed,dash phase=4pt] (n1.base)--+(225:0.8cm)
    node[below left=-2pt] {$X^{\dagger}$};
  \draw[dashed] (n1) arc (225:270:1cm) node (n2) {};
  \draw[dashed,double,dash phase=4pt]
    (n2.base)--+(270:0.8cm) node[below] {$F$};
  \draw[dashed] (n2) arc (270:315:1cm) node (n3) {};
  \draw[dashed,dash phase=4pt]
    (n3.base)--+(315:0.8cm) node[below right=-2pt] {$X^\dagger$};
  \draw[dashed,double] (n3) arc (315:360:1cm);
  \filldraw[black] (n1) circle (1pt) node[above right=-2pt]
    {$\lambda^\ast$};
  \filldraw[black] (n2) circle (1pt) node[above] {$\lambda$};
  \filldraw[black] (n3) circle (1pt) node[above left=-3pt]
    {$\lambda^\ast\hspace{-2pt}$};
  \draw[arr] (2,0)++(120:1.2cm) arc
    (120:150:1.2cm)--++(-122:1pt) node[midway,above=5pt] {$\,k$};
  \filldraw[cross,fill=white] (1,0) circle (4pt);
  \filldraw[cross,fill=white] (3,0) circle (4pt);
  \filldraw[black] (2,1) circle (1pt) node[below] {$\lambda^\ast$};
\end{tikzpicture}\,
+\hspace{-3pt}
\begin{tikzpicture}[>=latex',baseline=(vert_cent.base)]
  \node (vert_cent) {$\phantom{\cdot}$};
  \tikzstyle{arr}=[decoration={markings,mark=at position 1 with
    {\arrow[scale=1.5]{>}}}, postaction={decorate}];
  \draw[arr] (0,0)--(0.6,0) node[midway,above] {$p$};
  \draw[arr] (3.4,0)--(4,0) node[midway,above] {$p$};
  \draw[dashed] (2,0) ++(0:1cm) arc (0:90:1cm);
  \draw[dashed,dash phase=4pt] (2,1)--(2,1.8) node[above] {$X^\dagger$};
  \draw[dashed,double] (1,1) ++(0:1cm) arc (90:180:1cm);
  \draw[dashed] (2,0) ++(0:-1cm) arc (180:225:1cm) node (n1) {};
  \draw[dashed,double,dash phase=4pt] (n1.base)--+(225:0.8cm) node[below
    left=-2pt] {$F^{\dagger}$}; \draw[dashed] (n1) arc (225:270:1cm) node
    (n2) {};
  \draw[dashed,double,dash phase=4pt] (n2.base)--+(270:0.8cm)
    node[below] {$F$};
  \draw[dashed] (n2) arc (270:315:1cm) node (n3) {};
  \draw[dashed,dash phase=4pt] (n3.base)--+(315:0.8cm) node[below
    right=-2pt] {$X^\dagger$};
  \draw[dashed,double] (n3) arc (315:360:1cm);
  \filldraw[black] (n1) circle (1pt) node[above right=-2pt]
    {$\lambda^\ast$};
  \filldraw[black] (n2) circle (1pt) node[above]
    {$\lambda$};
  \filldraw[black] (n3) circle (1pt) node[above left=-3pt]
    {$\lambda^\ast\hspace{-2pt}$};
  \draw[arr] (2,0) ++(120:1.2cm) arc
    (120:150:1.2cm)--++(-122:1pt) node[midway,above=5pt] {$\,k$};
  \filldraw[cross,fill=white] (1,0) circle (4pt);
  \filldraw[cross,fill=white] (3,0) circle (4pt);
  \filldraw[black] (2,1) circle (1pt) node[below] {$\lambda^\ast$};
\end{tikzpicture}\,
+\hspace{-3pt}
\begin{tikzpicture}[>=latex',baseline=(vert_cent.base)]
  \node (vert_cent) {$\phantom{\cdot}$};
  \tikzstyle{arr}=[decoration={markings,mark=at position 1 with
    {\arrow[scale=1.5]{>}}}, postaction={decorate}];
  \draw[arr] (0,0)--(0.6,0) node[midway,above] {$p$};
  \draw[arr] (3.4,0)--(4,0) node[midway,above] {$p$};
  \draw[dashed,double] (2,0) ++(0:1cm) arc (0:90:1cm);
  \draw[dashed,dash phase=4pt] (2,1)--(2,1.8) node[above] {$X^\dagger$};
  \draw[dashed] (1,1) ++(0:1cm) arc (90:180:1cm);
  \draw[dashed,double] (2,0) ++(0:-1cm) arc (180:225:1cm) node (n1) {};
  \draw[dashed,dash phase=4pt] (n1.base)--+(225:0.8cm) node[below
    left=-2pt] {$X^{\dagger}$};
  \draw[dashed] (n1) arc (225:270:1cm) node (n2) {};
  \draw[dashed,double,dash phase=4pt] (n2.base)--+(270:0.8cm) node[below]
  {$F$};
  \draw[dashed] (n2) arc (270:315:1cm) node (n3) {};
  \draw[dashed,double,dash phase=4pt] (n3.base)--+(315:0.8cm)
    node[below right=-2pt] {$F^\dagger$}; \draw[dashed] (n3) arc
    (315:360:1cm);
  \filldraw[black] (n1) circle (1pt) node[above right=-2pt]
    {$\lambda^\ast$};
  \filldraw[black] (n2) circle (1pt) node[above] {$\lambda$};
  \filldraw[black] (n3) circle (1pt) node[above left=-3pt]
    {$\lambda^\ast\hspace{-2pt}$};
  \draw[arr] (2,0)++(120:1.2cm) arc (120:150:1.2cm)--++(-122:1pt)
    node[midway,above=5pt] {$\,k$};
  \filldraw[cross,fill=white] (1,0) circle (4pt);
  \filldraw[cross,fill=white] (3,0) circle (4pt);
  \filldraw[black] (2,1) circle (1pt) node[below] {$\lambda^\ast$};
\end{tikzpicture}\, ,}[]
from which we obtain
\eqn{\tilde{c}_{(X^\dagger)^2F^\dagger
F}(0)=-\frac{\lambda^\ast}{4\pi^2\lambda|X|^4},}[]
again using the dispersion method of section \ref{OPEmethod}. Thus, the
leading contribution to $B_\mu$ is
\eqn{B_\mu\approx-\frac18\frac{\lambda_u \lambda_d
\lambda^\ast}{16\pi^2\lambda}\frac{|F|^2}{X^2}.}[approxBmu]
This is 12.5\% of the answer in~\cite{Dvali:1996cu}. This underestimation
may be attributed to the cancelations that occur for lower-dimension
operators like the one in \BmuCancelation. Again, a resummation of the
contributions of the operators $Q^2(X^\dagger F^{\dagger} (X^\dagger X)^n
(F^{\dagger}F)^m)$, $m,n\ge0$, should yield the full one-loop result
of~\cite{Dvali:1996cu}.

For the calculation of $\delta A_{u,d}$ and $\delta m^2_{H_{u,d}}$, we will
first compute contributions to the OPEs $\Phi_1\Phi_2(x)\times
\Phi_1^\dagger\Phi_2^\dagger(0)$ and $\widetilde{\Phi}_1
\widetilde{\Phi}_2(x)\times\widetilde{\Phi}_1^\dagger
\widetilde{\Phi}_2^\dagger(0)$, and then act with $Q^2$ and $Q^2\bar{Q^2}$
according to \HiggsParams. The leading operator of the type $O_0$ (as in
(\ref{amh2OPE})) is $X^\dagger X$. Its Wilson coefficient at one loop is
given by
\eqn{-\frac{i}{2}\tilde{c}_{X^\dagger X}(s)=\hspace{-4pt}
\begin{tikzpicture}[>=latex',baseline=(vert_cent.base)]
  \node (vert_cent) {$\phantom{\cdot}$};
  \tikzstyle{arr}=[decoration={markings,mark=at position 1 with
    {\arrow[scale=1.5]{>}}}, postaction={decorate}];
  \draw[arr] (0,0)--(0.6,0) node[midway,above] {$p$};
  \draw[arr] (3.4,0)--(4,0) node[midway,above] {$p$};
  \draw[dashed,dash phase=-1pt] (2,0) ++(0:1cm) arc (0:180:1cm);
  \draw[dashed,dash phase=-1pt] (2,0) ++(0:-1cm) arc (180:360:1cm);
  \filldraw[cross,fill=white] (1,0) circle (4pt);
  \draw[arr] (2,0) ++(120:1.2cm) arc (120:150:1.2cm)--++(-122:1pt)
    node[midway,above=5pt] {$\,k$};
  \filldraw[cross,fill=white] (3,0) circle (4pt);
  \draw[dashed,dash phase=4pt] (2,1)--+(45:0.8cm) node[above right=-2pt]
    {$X^\dagger$};
  \draw[dashed,dash phase=4pt] (2,1)--+(135:0.8cm) node[above left=-2pt]
    {$X$};
  \filldraw[black] (2,1) circle (1pt) node[below] {$|\lambda|^2$};
\end{tikzpicture}\,
+\hspace{-2pt}
\begin{tikzpicture}[>=latex',baseline=(vert_cent.base)]
  \node (vert_cent) {$\phantom{\cdot}$};
  \tikzstyle{arr}=[decoration={markings,mark=at position 1 with
    {\arrow[scale=1.5]{>}}}, postaction={decorate}];
  \draw[arr] (0,0)--(0.6,0) node[midway,above] {$p$};
  \draw[arr] (3.4,0)--(4,0) node[midway,above] {$p$};
  \draw[dashed,dash phase=-1pt] (2,0) ++(0:1cm) arc (0:180:1cm);
  \draw[dashed,dash phase=-1pt] (2,0) ++(0:-1cm) arc (180:360:1cm);
  \draw[arr] (2,0) ++(120:1.2cm) arc (120:150:1.2cm)--++(-122:1pt)
    node[midway,above=5pt] {$\,k$};
  \filldraw[cross,fill=white] (1,0) circle (4pt);
  \filldraw[cross,fill=white] (3,0) circle (4pt);
  \draw[dashed,dash phase=4pt] (2,-1)--+(225:0.8cm) node[below left=-2pt]
    {$X$};
  \draw[dashed,dash phase=4pt] (2,-1)--+(315:0.8cm) node[below right=-4pt]
    {$X^\dagger$};
  \filldraw[black] (2,-1) circle (1pt)node[above] {$|\lambda|^2$};
\end{tikzpicture}\, ,}[]
from which we can compute
\eqn{\tilde{c}_{X^\dagger X}(0)=-\frac{1}{16\pi^2|X|^2}}[]
by the same dispersion methods as before. With this result we obtain
\eqn{\delta A_{u,d}\approx\frac12\frac{|\lambda_{u,d}|^2}{16\pi^2}
\frac{F^\dagger}{X^\dagger}, \qquad \delta m^2_{H_{u,d}}\approx\frac12
\frac{|\lambda_{u,d}|^2}{16\pi^2}\frac{|F|^2}{|X|^2}.}[]
For $\delta A_{u,d}$ this approximation gives 50\% of the full one-loop
result~\cite{Giudice:1997ni}, but for $\delta m_{H_{u,d}}^2$ we get an
answer that at first seems incompatible with the one-loop result.  Indeed,
due to an accidental cancelation the one-loop result for $\delta
m_{H_{u,d}}^2$ vanishes at leading order in $F/M^2$~\cite{Dvali:1996cu}.
One can also understand this result from arguments using analytic
continuation in superspace~\cite{Giudice:1997ni, Craig:2012xp}.  Our result
for $\delta m_{H_{u,d}}^2$ shows that the OPE, although very useful, can be
misleading in such approximate computations. Of course, if the full
computation within the OPE is performed, we should recover the full result
$\delta m_{H_{u,d}}^2=0$ at leading order. We will now show this
explicitly.

In order to carry out the analysis, we need to compute the Wilson
coefficient of the operator $(X^\dagger X)^n$, $n\geq1$, in the OPEs
$\Phi_1\Phi_2(x)\times \Phi_1^\dagger\Phi_2^\dagger(0)$ and
$\widetilde{\Phi}_1 \widetilde{\Phi}_2(x)\times\widetilde{\Phi}_1^\dagger
\widetilde{\Phi}_2^\dagger(0)$. We find
\eqn{\tilde{c}_{(X^\dagger X)^n}(s)=\frac{1}{4\pi^2}
\frac{\Gamma(2n-1)}{\Gamma(n) \Gamma(n+1)}\left(
\frac{|\lambda|^2}{s} \right)^n\ln\frac{-s}{\mu^2}+\cdots,}[]
and we can now use \integrate and \HiggsParams to obtain the full one-loop
result for both $\delta A_{u,d}$ and $\delta m_{H_{u,d}}^2$ at leading
order in $F/M^2$.  We find
\eqn{\delta A_{u,d}=\frac{|\lambda_{u,d}|^2}{8\pi^2}\left(\sum_{n=1}^\infty
\frac{n\Gamma(2n-1)}{2^{2n} \Gamma^2(n+1)}\right)\frac{F^\dagger}
{X^\dagger}=\frac{|\lambda_{u,d}|^2}{16\pi^2}\frac{F^\dagger}{X^\dagger}
}[Afull]
and
\eqn{\delta m_{H_{u,d}}^2=\frac{|\lambda_{u,d}|^2}{8\pi^2}
\left(\sum_{n=1}^\infty\frac{n^2\Gamma(2n-1)}{2^{2n} \Gamma^2(n+1)}\right)\frac{|F|^2}{|X|^2}.}[mHfull]
The sum in \Afull is convergent, and \Afull agrees precisely with the full
result (at one-loop and leading order in $F/M^2$) from Feynman diagram
computations as well as analytic continuation methods.

On the other hand, the sum in \mHfull is divergent, but it can be
regulated. The most divergent part in $\frac{n^2\Gamma(2n-1)}{2^{2n}
\Gamma^2(n+1)}= \frac{1}{16\sqrt{\pi}} \frac{\Gamma(n-1/2)}{\Gamma(n)}$ is
$\frac{1}{16\sqrt{\pi}} \frac{1}{\sqrt{n}}$, and we know from
$\zeta$-function regularization that
\eqn{\sum_{n=1}^\infty\frac{1}{\sqrt{n}}=\zeta(1/2).}[DivPart]
Furthermore, subtracting the most divergent part from the full sum and
regularizing with a $z^n$ we find
\eqn{\sum_{n=1}^\infty\left(\frac{\Gamma(n-1/2)}{\Gamma(n)}
-\frac{1}{\sqrt{n}}\right)z^n =
\frac{\sqrt{\pi}z}{\sqrt{1-z}}-\Li_{1/2}(z),}[ConvPart]
where $\Li_n(z)=\sum_{k=1}^\infty\frac{z^k}{k^n}$ is the polylogarithm
function. But \ConvPart has a well defined limit as $z\to 1$:
\eqn{\lim_{z\to1}\left(\frac{\sqrt{\pi}z}{\sqrt{1-z}}-\Li_{1/2}(z)\right)=-
\zeta(1/2).}[ConvPartLimit]
Adding \DivPart and \ConvPart and using \ConvPartLimit we thus obtain
$\delta m_{H_{u,d}}^2=0$ at leading order and one loop. This agrees
with~\cite{Dvali:1996cu} and further elucidates the power of the OPE in
this context.

Regarding the parameters in $V^{(\text{other})}_{\text{Higgs}}$, it is
straightforward to see using our methods that they are suppressed relative
to those in $V^{(\text{soft})}_{\text{Higgs}}$.  We will not show this
explicitly here.

\subsec{A comment on the OPE with unrenormalized operators}
Before we conclude this section, let us elaborate on a feature that may be
of interest. Consider, as we have so far, the OPE in momentum space,
\eqn{i\int\dfour x\,e^{-ip\cdot x}\CO_i(x)\CO_j(0) = \sum_{k}
\tilde{c}_{\smash{ij}}^{\phantom{ij\!}k}(p^2)\CO_k (0).}[OPEMomSp]
In computations of the coefficients
$\tilde{c}_{\smash{ij}}^{\phantom{ij\!}k}$, we found that the introduction
of an arbitrary normalization point $\mu$ (not to be confused with the
$\mu$ parameter) was necessary in order to avoid IR divergences, as in
$\tilde{c}_{X^\dagger F^\dagger}$ in \muWilCoeffOne for example. In other
words, we found that $\tilde{c}_{\smash{ij}}^{\phantom{ij\!}k}$ were
actually functions of $p^2$ and $\mu^2$. This is a known consequence of the
splitting of UV and IR physics inherent in the OPE~\cite{Novikov:1984rf}.
Of course the $\mu$-dependence is also there in the right-hand side of the
OPE when renormalized operators are used.

Now, note that if we use bare operators in the left-hand side of the OPE,
then there is no $\mu$-dependence in the corresponding (unrenormalized)
expectation value of the operator product. Such dependence, therefore, has
to disappear from the right-hand side of the OPE as well. To see this
cancellation of the renormalization-point dependence, one needs to consider
operator mixing and the $\mu$-dependence of operators in the right-hand
side of the OPE.

To illustrate this, let us consider the operators $X^\dagger F^\dagger$ and
$X^\dagger\Phi\widetilde{\Phi}\equiv X^\dagger(\Phi_1\widetilde{\Phi}_1
+\Phi_2\widetilde{\Phi}_2)$. Clearly, both  of them appear in the
right-hand side of the OPE $Q^\alpha(\Phi_1\Phi_2(x))\times Q_\alpha
(\widetilde{\Phi}_1\widetilde{\Phi}_2(0))$, and they mix under
renormalization since they have the same classical scaling dimension and
quantum numbers. As they have appeared up to now, these operators have an
implicit $\mu$-dependence. What we have neglected up to now, however, is
that these operators mix under renormalization. At one loop, and without
taking into account wavefunction renormalization (which is unnecessary for
the point being made here), we have
\eqn{(X^\dagger\Phi\widetilde{\Phi}(0))_{\text{Bare}}
=X^\dagger\Phi\widetilde{\Phi}(0)
+\frac{\lambda^\ast}{8\pi^2}\ln\frac{\Lambda^2}{\mu^2}X^\dagger
F^\dagger(0),}[OpMix]
where $\Lambda$ is the UV cutoff, as computed from the graph
\vspace{-5pt}
\eqn{
\begin{tikzpicture}[>=latex',baseline=(vert_cent.base)]
  \node (vert_cent) {$\phantom{\cdot}$};
  \tikzstyle{arr}=[decoration={markings,mark=at position 1 with
    {\arrow[scale=1.5]{>}}}, postaction={decorate}];
  \draw[dashed,dash phase=-1pt] (2,0) ++(0:1cm) arc (0:180:1cm);
  \draw[dashed,dash phase=-1pt] (2,0) ++(0:-1cm) arc (180:360:1cm);
  \draw[dashed,dash phase=1pt] (1,0)--+(180:0.93cm) node[left=-2pt]
    {$X^\dagger$};
  \filldraw[cross,fill=white] (1,0) circle (4pt);
  \draw[arr] (2,0) ++(70:1.2cm) arc (70:110:1.2cm)--++(197:2pt)
    node[midway,above right=2pt] {$\,\,k$};
  \draw[dashed,dash phase=4pt,double] (3,0)--+(0:0.8cm) node[right]
    {$F^\dagger$};
  \filldraw[black] (3,0) circle (1pt) node[left] {$\lambda^\ast$};
\end{tikzpicture},}[]
which is both UV and IR divergent and so the integration over virtual
momenta is cut-off at $\Lambda$ from above and at $\mu$ from below.
Although the vev of $X^\dagger\Phi\widetilde{\Phi}$ is finite, that of
$(X^\dagger\Phi\widetilde{\Phi})_{\text{Bare}}$ is infinite. Now, since the
tree-level Wilson coefficient of $X^\dagger\Phi\widetilde{\Phi}$ in the OPE
$Q^\alpha(\Phi_1\Phi_2(x))\times Q_\alpha
(\widetilde{\Phi}_1\widetilde{\Phi}_2(0))$ is given by
\eqn{-i\tilde{c}_{X^\dagger\Phi\widetilde{\Phi}}(s)=-2\times\!\!
\begin{tikzpicture}[>=latex',baseline=(vert_cent.base)]
  \node (vert_cent) {$\phantom{\cdot}$};
  \tikzstyle{arr}=[decoration={markings,mark=at position 1 with
    {\arrow[scale=1.5]{>}}}, postaction={decorate}];
  \draw[arr] (0,0)--(0.6,0) node[midway,above] {$p$};
  \draw[arr] (3.4,0)--(4,0) node[midway,above] {$p$};
  \draw (1,0)--(3,0);
  \draw[dashed,dash phase=1pt] (1,0)--+(270:0.93cm) node[below]
    {$\Phi_{1,2}$};
  \draw[dashed,dash phase=1pt] (3,0)--+(270:0.93cm) node[below=-3pt]
    {$\widetilde{\Phi}_{1,2}$};
  \filldraw[cross,fill=white] (1,0) circle (4pt);
  \filldraw[cross,fill=white] (3,0) circle (4pt);
  \draw[dashed,dash phase=4pt] (2,0)--+(90:0.8cm) node[above]
    {$X^\dagger$};
  \filldraw[black] (2,0) circle (1pt) node[below] {$\lambda^\ast$};
\end{tikzpicture}\,
=-\frac{4i\lambda^\ast}{s},}[]
we see that the $\mu$ in \muWilCoeffOne is indeed replaced by $\Lambda$ via
operator mixing between $X^\dagger F^\dagger$ and
$X^\dagger\Phi\widetilde{\Phi}$, at this order in $\lambda$, when
$X^\dagger\Phi\widetilde{\Phi}$ is substituted by
$(X^\dagger\Phi\widetilde{\Phi})_{\text{Bare}}$ via \OpMix.  Similar
replacements $\mu\to\Lambda$ in physical logarithms of Wilson coefficients
will occur order by order in perturbation theory. All other explicit
$\mu$-dependence in the $\xi$-expansion of \muWilCoeff, which is
regularization-scheme dependent, will similarly disappear when the
right-hand side of the OPE is written in terms of the bare operators.

\newsec{Comments on phenomenology}[pheno]
From the general results obtained in section \ref{computations}, we can
make various comments about possible phenomenological consequences. The
generality of the OPE formalism has a great advantage in that it provides a
powerful organizing principle. It not only leads to an understanding of
known phenomenological problems from a different (possibly deeper)
perspective, but also provides a unifying framework for describing the
various proposals for these problems.

To see how this works in more detail, let us look at the
$\mu/B_\mu$~\cite{Dvali:1996cu} and $A/m_{H}^2$~\cite{Craig:2012xp}
problems, which in essence are manifestations of the problem of achieving
phenomenologically viable electroweak symmetry breaking (EWSB).  These
problems arise in many models with Higgs-messenger interactions trying to
generate the $\mu$ and $A$ terms, as such models also tend to produce
$B_\mu$ and $m_H^2$ terms which are too large to be viable for
EWSB.\footnote{The $A/m_H^2$ problem is absent in ``Minimal Gauge
Mediation'' (MGM) with a single spurion, since in this case $m_{H}^2$
vanishes at one-loop at leading order in $F/M$ (this can be seen explicitly
in the model of section \ref{toy}). But even in this case, a ``little
$A/m_H^2$ problem'' remains if one wants large
$A$-terms~\cite{Craig:2013wga}.} The results in section \ref{computations}
allow us to understand these problems from a different perspective. For
instance, we saw in (\ref{muterm}) that operators $O_{3,0}$,
$Q^2\bar{Q}^2O_{3,0}$ and $Q\sigma_{\mu}\bar{Q}O_{3,1}^\mu$ contribute to
$\mu$, in contrast to only $Q^2O_{3,0}$ contributing to $B_\mu$ (see
(\ref{Bmuterm})). In simple models, as in the weakly coupled example in
section \ref{toy}, only $O_{3,0}$ contributes to $\mu$\footnote{In this
model, operators of type $Q^2\bar{Q}^2(O_{3,0;i})$ can always be written as
$O_{3,0;j}$ for some $i,j$.} while $Q^2\,O_{3,0}$ contributes to $B_\mu$.
Now, since it is assumed that $\mu$ is forbidden in the supersymmetric
limit, it usually arises at order $F$, i.e.\ $\langle O_{3,0}\rangle \propto
F$. Then, if the supersymmetry breaking dynamics is essentially trivial (as
in a spurion model), one obtains parametrically $\langle Q^2O_{3,0}\rangle
\propto F\langle O_{3,0}\rangle \propto F^2$.  Finally, in weakly coupled
models $B_\mu$ is typically generated at the same loop order as that of
$\mu$, so that the Wilson coefficients ${\hat{c}}_{B_\mu}\simeq
{\hat{c}}_{\mu} \sim 1/16\pi^2$.  This gives rise to the well-known
problematic relation
\eqn{\frac{B_\mu}{|\mu|^2} = \frac{{\hat{c}}_{B_\mu}\langle
Q^2O_{3,0}\rangle}{|{\hat{c}}_{\mu}|^2\,|\langle O_{3,0}\rangle|^2} \sim
16\pi^2.}[bmuproblem]
Similar statements can be made for the $A/m_H^2$ problem.

The OPE results suggest possible ways of addressing these issues. We will
consider the case when $\langle O_{3,0}\rangle \neq 0$, which is
generically at order $F$ (since $\mu$ is assumed to vanish in the
supersymmetric limit). In this case, if $Q\sigma_{\mu} \bar{Q}O_{3,1}^\mu$
gets a vev as well, then that vev is of order $F^2$ as mentioned in section
\ref{computations}. This is because $\langle
Q\sigma_{\mu}\bar{Q}O_{3,1}^\mu\rangle$ is of the form
$\langle\{Q,[\bar{Q},(\cdot)]\}\rangle$, and all such correlation functions
must start at order $F^2$~\cite{Komargodski:2008ax}. Thus, this operator
(and also $Q^2\bar{Q}^2O_{3,0}$) will only have a subleading effect on
$\mu$.  In this case, it is also easy to see that the wrong-Higgs
trilinears $a'_{u,d}$ will be suppressed compared to $\mu$ and other soft
terms expected to be generated at order $F$. In this situation, one can
imagine two possible ways of solving the problem of achieving viable EWSB.
Let us now discuss each of these from the perspective of the OPE results
obtained.

\subsec{EWSB with \texorpdfstring{$B_\mu/\mu^2 \lesssim
1$}{Bmu/musq<1} and \texorpdfstring{$m_{H_{u,d}}^2/A_{u,d}^2 \lesssim
1$}{mHsq/Asq<1} }\label{confess}
From the discussion in the previous sections, we know that the $\mu$
parameter is given predominantly by $\langle O_{3,0}\rangle$, while the
$B_\mu$ parameter is given by $\langle Q^2\,O_{3,0}\rangle$. Similarly,
$A_{u,d}$ is given by $\langle Q^2O^{u,d}_{0}\rangle$ while
$m_{H_{u,d}}^2$ is given by $\langle Q^2\bar{Q}^2O^{u,d}_{0}\rangle$.
An interesting feature of the OPE results is that the operators $O_{3,0}$
and $O^{u,d}_{0}$ are completely independent on each other in general.
Also, for a strongly coupled extra sector, the vevs of operators are
determined by strong infrared dynamics and cannot be computed in general.
Finally, for a strongly coupled sector with nontrivial supersymmetry
breaking dynamics (that is not captured by a single spurion), na\"ive
parametrics such as $\langle Q^2O_{3,0}\rangle \propto F\langle
O_{3,0}\rangle \propto F^2$, giving rise to the problematic relation
(\ref{bmuproblem}), need not hold. Thus, with nontrivial supersymmetry
breaking dynamics, it is possible to have viable EWSB with  $B_\mu/\mu^2
\lesssim 1$ and $m_{H_{u,d}}^2/A_{u,d}^2 \lesssim 1$, as long as
\eqn{\frac{\langle Q^2O_{3,0}\rangle}{|\langle O_{3,0}\rangle|^2}\simeq
\frac{|{\hat{c}}_{\mu}|^2}{{\hat{c}}_{B_\mu}},\qquad
\frac{\langle Q^2\bar{Q}^2O^{u,d}_{0}\rangle}{|\langle
Q^2\,O^{u,d}_{0}\rangle|^2}\simeq \frac{|{\hat{c}}_{A_{u,d}}|^2}
{{\hat{c}}_{m_{H_{u,d}}^2}},}[parametricsReqd]
respectively. The relations \BmumuWilson and \mHAWilson can also be used in
\parametricsReqd. Note that the ratio $B_\mu/\mu^2$ and
$m_{H_{u,d}}^2/A_{u,d}^2$ will be different in general as they are
controlled by the dynamics of different operators $O_{3,0}$ and
$O^{u,d}_0$.  Finally, it is worth remembering that the parameters in the
Higgs Lagrangian computed above correspond to parameters at the scale $M$.
In order for the mechanism described above to give viable EWSB, the RG
evolution of Wilson coefficients of operators to scale
$\sqrt{F}$,\footnote{We have assumed $F < M^2$.} where the supersymmetry
breaking sector fully decouples, should be negligible.  For example, this
could happen if all states in the extra sector have mass of order $M$ (with
supersymmetry breaking splittings $\sim \sqrt{F}\ll M$), and the
renormalization from $M$ to $\sqrt{F}$ from the supersymmetry breaking
sector is trivial.\footnote{This has been effectively assumed in the
branch-cut structure in section \ref{OPEmethod}. Also, note that
renormalization of the parameters from \emph{visible} sector states is
still present below the scale $M$.}

There exists another possibility,  however, \emph{viz.}\ that the relations
between the Higgs parameters at scale $M$ are given by the na\"ive
expressions such as (\ref{bmuproblem}), but there is strong renormalization
of Wilson coefficients of operators relevant to $\mu/B_\mu$ and $A/m_H^2$
between $M$ and $\sqrt{F}$. This mechanism, which has been known for quite
some time, is that of conformal sequestering~\cite{Luty:2001jh,
Luty:2001zv, Dine:2004dv, Schmaltz:2006qs, Murayama:2007ge, Roy:2007nz}.
Within the OPE formalism, this mechanism can be understood as follows. In
order to describe the flow from $M$ to $\sqrt{F}$, one has to integrate out
states with mass of order $M$, and write down an effective theory in terms
of the light degrees of freedom with some coefficients. These coefficients
can be obtained by matching to the full theory at the scale $M$ and must
then be RG-evolved to lower scales. If this resulting effective theory is
itself an approximately superconformal field theory (\emph{different} from
the UV superconformal theory in general), then the coefficients of the
operators for the Higgs parameters in the effective theory will receive a
power-law running.

In general, the matching on to an effective theory at the scale $\simeq M$
is model-dependent. However, one framework, in which the matching is
tractable, is that of the case of GMHM~\cite{Craig:2013wga}, in which it is
assumed that the extra sector factorizes into a messenger sector and a
supersymmetry breaking (hidden) sector, coupled by a small dimensionless
coupling $\kappa$ of the form
\eqna{W \supset \kappa\Lambda^{-(\Delta_{\CO_h}
+\Delta_{\CO_m}-3)}{\CO}_h\CO_m.}[]
In this case it is possible to integrate out operators $\CO_m$ and consider
the effective theory with $\CO_h, \CO_h^{\dag}$ coupled to the Higgs
fields.  Then, the usual expressions for conformal sequestering of
coefficients of the relevant operators can be derived~\cite{Craig:2013wga}.
To complete this section, it is worth noting that the case of extreme
sequestering is disfavored both phenomenologically~\cite{Perez:2008ng,
Asano:2008qc} and theoretically~\cite{Poland:2011ey}; see detailed
discussion in~\cite{Knapen:2013zla}.  However,~\cite{Knapen:2013zla} shows
that it is possible to obtain viable electroweak symmetry breaking with
mild sequestering: $B_\mu/\mu^2\lesssim 1$ and $m_{H_u}^2/A_u^2\lesssim 1$,
and suitable Wilson coefficients.

\subsec{EWSB with ``lopsided'' gauge mediation}[lopsided]
Another mechanism for obtaining viable electroweak symmetry breaking is via
what is known as ``lopsided'' gauge mediation~\cite{DeSimone:2011va}. In
this class of models, viable EWSB is possible even with $B_\mu \gg \mu^2 $
as long as one also has $m_{H_u}^2 \ll B_\mu \ll m_{H_d}^2$. It was argued
in~\cite{Csaki:2008sr} that this can be arranged if the couplings
$\lambda_u, \lambda_d$ satisfy $\lambda_u \ll \lambda_d$ (for the analysis
in this paper to be valid, both couplings should still be perturbative).
Furthermore, it was shown in~\cite{SchaferNameki:2010iz} that such a setup
can be naturally realized within a SQCD-like supersymmetry breaking sector
in which one of the Higgses ($H_d$) in the MSSM mixes strongly with the
supersymmetry-breaking sector and is composite. Therefore, the Higgs sector
in this case is ``hybrid,'' with $H_u$ being an elementary field while
$H_d$ being composite. Note that in this case the RG-evolution of the
Wilson coefficients of the various Higgs parameters from $M$ to $\sqrt{F}$
is negligible.

The OPE results obtained in this paper suggest more general possibilities
for obtaining the pattern $B_\mu \gg \mu^2 $ and $m_{H_u}^2 \ll B_\mu \ll
m_{H_d}^2$, \emph{in addition} to the $\lambda_u \ll \lambda_d$ possibility
above. This is because of the same reason as mentioned earlier: the vevs of
operators appearing in the OPE for the various Higgs parameters are
determined by infrared dynamics in general, and the na\"ive parametric
relations between different Higgs parameters may not hold. For example,
even if $\lambda_u$ and $\lambda_d$ are comparable to each other, it is
possible to obtain $m_{H_u}^2 \ll B_\mu \ll m_{H_d}^2$ if
\eqn{{\hat{c}}_{m_{H_u}^2} \langle Q^2\bar{Q}^2O^u_0\rangle \ll
{\hat{c}}_{B_\mu}\langle Q^2O_{3,0}\rangle \ll {\hat{c}}_{m_{H_d}^2}
\langle Q^2\bar{Q}^2O^d_0\rangle.}[lopsided-eq]
In addition, it is also possible to have $|A_u|^2 > m_{H_u}^2$ if
$\frac{\langle Q^2\bar{Q}^2\,O^u_0\rangle}{|\langle Q^2\,O^u_0\rangle|^2} <
\frac{|{\hat{c}}_{A_u}|^2}{{\hat{c}}_{m_{H_u}^2}}$ so that the Higgs
quartic coupling gets nontrivial radiative contributions from visible
sector superpartners such as stops. Note that the coupling (\ref{W}) will
also provide a source of extra contributions to the Higgs quartic
couplings, and it would be very interesting to compute the general form of
these corrections.

\newsec{Summary and future directions}[discuss]
In this work we have developed a systematic and general formalism to
compute the parameters in the effective Higgs Lagrangian to quadratic
order, in a broad class of supersymmetric frameworks in which the Higgs
fields in the visible sector couple to another sector via the
superpotential \W. It is assumed that the additional sector is
superconformal in the UV but develops a mass gap $\sim M$ and supersymmetry
breaking splitting $\sim \sqrt{F}$ with $F/M$ not far from the TeV scale.
The primary technique used to compute the Higgs parameters is that of the
operator product expansion (OPE). The results obtained within this
formalism are completely general within the class of frameworks described
above, and can be applied even when the additional sector is strongly
coupled. The formalism provides a deeper insight into problems affecting
simple models of supersymmetry breaking and mediation, and provides new
possibilities for solutions. The underlying reason for the existence of
these new possibilities which have not been considered before, is the fact
that OPE methods imply that different types of operators contribute to
different Higgs parameters in general.  Furthermore, since the vevs of
these operators are determined by infrared dynamics, simple parametric
relations between different Higgs parameters, which hold in weakly coupled
or spurion based models, may not hold in general.

There are a few interesting directions that are worth exploring in the
future.  One interesting direction would be to construct realistic models
of dynamical supersymmetry breaking and mediation, and apply OPE techniques
to explicitly compute the Wilson coefficients and the relevant Higgs
parameters. It would also be worth exploring the computation of the quartic
terms in the effective Higgs Lagrangian, due to the presence of the
superpotential couplings (\ref{W}). This would be crucial for computing the
physical Higgs boson masses and mixing angles, and as such is directly
relevant for phenomenology.  In particular, it is straightforward to see
that the contribution to the Higgs quartic couplings, for example the
coefficient $\lambda_{H_u}$ in $\lambda_{H_u}|H_u|^4$, due to the couplings
(\ref{W}), will be determined by a four-point function of the form
\eqn{|\lambda_u|^4\,\left\langle \int
\dfour y\,\dfour z\,\dfour w\,Q^2O_u(0)\bar{Q}^2O_u^{\dag}(y)Q^2
O_u(z)\bar{Q}^2O_u^{\dag}(w)\right\rangle.}[]
Within the OPE formalism described above, a possible way to compute such
quartic couplings is by expanding these four-point functions in conformal
blocks~\cite{Dolan:2000ut}, or their supersymmetric extensions, the
superconformal blocks for ${\cal N}=1$ supersymmetry~\cite{Poland:2010wg}.
It would be interesting to develop this approach to the quartic couplings
in greater detail.

Another interesting direction to explore would be to apply similar OPE
methods to compute terms in the effective Higgs Lagrangian for the case
when the Higgs fields couple to a SM gauge-singlet $\mathcal{S}$, via
$\mathcal{S}H_u H_d$ in the superpotential. Although such operators have
been studied extensively in the literature (mostly in weakly-coupled
settings), OPE results may shed an interesting light on the physics in more
general cases that would be hard to see otherwise.

Finally, it would be interesting to extend our analysis of Higgs parameters
to include the possible contributions of poles appearing in the two-point
functions of the additional sector, similar to what was done for current
correlators in~\cite{Intriligator:2010be, Buican:2009vv, Kitano:2011fk,
McGarrie:2012ks, McGarrie:2012fi}.  In this case one could derive sum-rules
relating the contributions of these poles to the Higgs parameters,
obtaining results that are applicable to an even more general class of
models. We hope to study these and related directions in the near future.

\acknowledgments{We would like to thank Tom Appelquist, Andrew Cohen,
George Fleming, Walter Goldberger, and Martin Schmaltz for helpful
discussions. We also thank Jeff Fortin and Ken Intriligator for useful
discussions and comments on the manuscript. This work is supported in part
by DOE grant DE-FG-02-92ER40704.}

\begin{appendices}

\newsec{Two-point functions of superconformal and conformal
primaries}[technicalTwoP]
As we already saw in \TwoPoints, in an $\mathcal{N}=1$ superconformal
theory the two-point function of an operator $\CO^I$ with its conjugate
${\bar{\CO}}^{\bar{I}}$ is given by
\eqn{\langle\CO^I(z_1){\bar{\CO}}^{\bar{I}}(z_2)\rangle =
C_\CO\frac{\CI^{I\bar{I}}(x_{1\bar{2}},x_{\bar{1}2})}
{\osbx{\bar{1}2}{2\bar{q}}\osbx{\bar{2}1}{2q}}.}[]
Now, one can carry out the $\theta$-expansion of both sides of \TwoPoints
and match the various powers of $\theta_{1,2},\bar{\theta}_{1,2}$ between
the two sides to read off two-point functions of the various components of
$\CO^I$. This achieves a projection of the superconformal two-point
function to the conformal subgroup. Nevertheless, this projection is
contaminated by the presence of conformal descendants in the
$\theta$-expansion of $\CO^I$. This contamination has to be removed in
order to obtain a projection to conformal primaries.

To illustrate these points, let us work out explicitly the case of a
general scalar $\mathcal{N}=1$ superfield operator $\CO$. In this case
\TwoPoints becomes
\eqn{\langle\CO(z_1)\bar{\CO}(z_2)\rangle =
\frac{C_\CO}{\osbx{\bar{1}2}{2\bar{q}}\osbx{\bar{2}1}{2q}}.}[ScalarTP]
Now, the Baker--Campbell--Hausdorff formula and the supersymmetry algebra
imply that
\eqn{e^{i\theta Q+i\bar{\theta}\bar{Q}}=e^{i\theta
Q}e^{i\bar{\theta}\bar{Q}}e^{\theta P\cdot\sigma\bar{\theta}},}[]
and expanding the exponentials it is straightforward to evaluate
\eqna{e^{i\theta Q+i\bar{\theta}\bar{Q}}&=1+i\theta Q+i\bar{\theta}\bar{Q}
+\tfrac12\theta\sigma^\mu\bar{\theta}(Q\sigma_\mu\bar{Q}+2P_\mu)
+\tfrac14\theta^2Q^2+\tfrac14\bar{\theta}^{\hspace{0.5pt}2}\bar{Q}^2\\
&\quad-\tfrac{i}{4}\theta^2\bar{\theta}^{\dot{\alpha}}(Q^2
\bar{Q}_{\dot{\alpha}}-2Q^\alpha\sigma^\mu_{\alpha\dot{\alpha}}P_\mu)
+\tfrac{i}{4}\bar{\theta}^{\hspace{0.5pt}2}\theta^\alpha(\bar{Q}^2Q_\alpha
+2\sigma^\mu_{\alpha\dot{\alpha}}\bar{Q}^{\dot{\alpha}}P_\mu)\\
&\quad+\tfrac{1}{16}\theta^2\bar{\theta}^{\hspace{0.5pt}2}(Q^2\bar{Q}^2-4P^2
-4Q\sigma^\mu\bar{Q}P_\mu).}[exponExp]
This implies that $\CO(z)\equiv e^{i\theta Q+i\bar{\theta}\bar{Q}}O(x)$ can
be expanded as
\eqna{\CO(z)&=O+i\theta QO+i\bar{\theta}\bar{Q}O
+\tfrac12\theta\sigma^\mu\bar{\theta}([Q\sigma_\mu\bar{Q}O]_p+c_1P_\mu O)
+\tfrac14\theta^2Q^2O+\tfrac14\bar{\theta}^{\hspace{0.5pt}2}\bar{Q}^2O\\
&\quad-\tfrac{i}{4}\theta^2\bar{\theta}^{\dot{\alpha}}([Q^2
\bar{Q}_{\dot{\alpha}}O]_p-
c_2\sigma^\mu_{\alpha\dot{\alpha}}P_\mu Q^\alpha O)
+\tfrac{i}{4}\bar{\theta}^{\hspace{0.5pt}2}\theta^\alpha
([\bar{Q}^2Q_\alpha O]_p
-c_3\sigma^\mu_{\alpha\dot{\alpha}}P_\mu\bar{Q}^{\dot{\alpha}}O)\\
&\quad+\tfrac{1}{16}\theta^2\bar{\theta}^{\hspace{0.5pt}2}
([Q^2\bar{Q}^2 O]_p -c_4P^2 O
-c_5P_\mu [Q\sigma^\mu\bar{Q} O]_p),}[SuperFieldExp]
where $[\,\cdot\,]_p$ denotes a conformal primary operator, and
$c_{1,\ldots,5}$ are coefficients we need to evaluate, and that will allow
us to see which combinations of $Q\sigma_\mu\bar{Q}O$ and $P_\mu O$,
$\bar{Q}^2Q_\alpha O$ and $\sigma^\mu_{\alpha\dot{\alpha}}P_\mu
\bar{Q}^{\dot{\alpha}}O$, $Q^2\bar{Q}_{\dot{\alpha}}O$ and
$\sigma^{\mu}_{\alpha\dot{\alpha}}P_\mu Q^\alpha O$, and $Q^2\bar{Q}^2O$,
$P^2 O$ and $P_\mu[Q\sigma^\mu\bar{Q}]_p$ are conformal primaries. Note
that some components in the expansion are already conformal primaries; for
example $[Q^2O]_p=Q^2O$.

The coefficients $c_{1,\ldots,5}$ can be evaluated by expanding both sides
of \ScalarTP using \SuperFieldExp on the left-hand side. For example, from
the $\theta_1\bar{\theta}_1$ component of $\langle\CO(z_1)
\bar{\CO}(z_2)\rangle$ we can find
\eqn{c_1=2\frac{q-\bar{q}}{q+\bar{q}},}[]
and comparing \SuperFieldExp with \exponExp we find that
\eqn{[Q\sigma_\mu\bar{Q}O]_p=Q\sigma_\mu\bar{Q}O
+4\frac{\bar{q}}{q+\bar{q}}P_\mu O,}[]
which, as expected, is zero for $q=0$ or $\bar{q}=0$. With our result for
$c_1$ we can use the $\theta_1 \bar{\theta}_1 \theta_2 \bar{\theta}_2$ part
of \ScalarTP to compute
\eqn{\langle[Q\sigma^\mu\bar{Q}O(x)]_p[Q\sigma^\nu\bar{Q}O(0)]_p^\dagger
\rangle = 2^5C_\CO\frac{q\bar{q}(q+\bar{q}+1)}{q+\bar{q}}\frac{I^{\mu\nu}(x)}
{x^{2(q+\bar{q}+1)}},\qquad I^{\mu\nu}(x)=\eta^{\mu\nu}-2\frac{x^\mu
x^\nu}{x^2},}[VecTP]
which has the required form, $\sim I^{\mu\nu}$, dictated by
conformal invariance, and the correct behavior for chiral
($\bar{D}_{\dot{\alpha}}\Phi=0$, $\bar{q}=0$) and anti-chiral
($D_{\alpha}\Phi=0$, $q=0$) superfields. For a linear superfield
($D^2\mathcal{J}=\bar{D}^2\mathcal{J}=0$, $q=\bar{q}=1$), whose
$\theta\bar{\theta}$ component is a conserved vector current, we also get a
consistency check by the fact that $\partial_\mu\langle[Q\sigma^\mu
\bar{Q}O(x)]_p[Q\sigma^\nu\bar{Q}O(0)]_p^\dagger\rangle$ correctly vanishes
for $q=\bar{q}=1$.

We can also evaluate
\eqn{c_2=2\frac{q-\bar{q}-1}{q+\bar{q}-1},\qquad
c_3=2\frac{q-\bar{q}+1}{q+\bar{q}-1},}[]
and thus obtain
\twoseqn{
\langle[Q^2\bar{Q}_{\dot{\alpha}}O(x)]_p [\bar{Q}^2Q_\alpha O^\dagger(0)]_p
\rangle&=-2^8iC_\CO\frac{q\bar{q}(q-1)(q+\bar{q}+1)}{q+\bar{q}-1}
\frac{\text{x}_{\alpha\dot{\alpha}}}{x^{2(q+\bar{q}+2)}},}[QQbsqTP]{
\langle[\bar{Q}^2Q_\alpha O(x)]_p[Q^2\bar{Q}_{\dot{\alpha}}
O^\dagger(0)]_p\rangle&=-2^8iC_\CO\frac{q\bar{q}(\bar{q}-1)(q+\bar{q}+1)}
{q+\bar{q}-1}\frac{\text{x}_{\alpha\dot{\alpha}}}{x^{2(q+\bar{q}+2)}},
}[QbQsqTP][QQbsqorQbQsqTP]
where
\twoseqn{[Q^2\bar{Q}_{\dot\alpha}O]_p&=Q^2\bar{Q}_{\dot\alpha}O
-4\frac{\bar{q}}{q+\bar{q}-1}\sigma^\mu_{\alpha\dot{\alpha}}P_\mu
Q^{\alpha}O,}[QsqQbprim]{[\bar{Q}^2Q_\alpha O]_p&=\bar{Q}^2Q_\alpha O
+4\frac{q}{q+\bar{q}-1}\sigma^\mu_{\alpha\dot{\alpha}}P_\mu
\bar{Q}^{\dot\alpha} O.}[QbsqQprim]
It is easy to see that \QsqQbprim becomes zero for $q=0$, $\bar{q}=0$ or
$q=1$, and correspondingly \QbsqQprim becomes zero when $\bar{q}=0$, $q=0$
or $\bar{q}=1$.  Expressions \QQbsqorQbQsqTP correctly become zero if $\CO$
is a chiral, antichiral, or linear superfield. Furthermore, as expected,
\QQbsqTP and \QbQsqTP become zero for a scalar superfield $\mathcal{S}$
satisfying $D^2 \mathcal{S}=0$ or $\bar{D}^2 \mathcal{S}=0$
respectively.\foot{A superfield $\mathcal{S}$ satisfying $\bar{D}^2
\mathcal{S}=0$ has twelve fermionic and twelve bosonic degrees of freedom
and is called a nonminimal (or complex) scalar superfield. Such a
superfield is a superconformal quasi-primary, and has $\bar{q}=1$ and
$q>1$.  Correspondingly, a superfield $\mathcal{S}$ satisfying
$D^2\mathcal{S}=0$ has $q=1$ and $\bar{q}>1$.}

With a little bit more work we can compute
\eqn{c_4=4\frac{(q-\bar{q})^2-q-\bar{q}}{(q+\bar{q})(q+\bar{q}-1)},\qquad
c_5=4\frac{q-\bar{q}}{q+\bar{q}-2},}[]
which allow us to find
\eqn{\langle[Q^2\bar{Q}^2 O(x)]_p[Q^2\bar{Q}^2 O(0)]_p^\dagger\rangle =
2^{12}C_\CO\frac{q\bar{q}(q-1)(\bar{q}-1)(q+\bar{q})(q+\bar{q}+1)}
{(q+\bar{q}-1)(q+\bar{q}-2)}\frac{1}{x^{2(q+\bar{q}+2)}},}[QsqQbsqTP]
where
\eqn{[Q^2\bar{Q}^2O]_p=Q^2\bar{Q}^2O-2^4\frac{\bar{q}(\bar{q}-1)}
{(q+\bar{q}-1)(q+\bar{q}-2)}P^2O -
8\frac{\bar{q}-1}{q+\bar{q}-2}P_\mu Q\sigma^\mu\bar{Q}O,}[]
which correctly goes to zero for $\bar{q}=0$ or $\bar{q}=1$, as well as
$q=0$ or $q=1$. As expected, \QsqQbsqTP becomes zero for a chiral and an
antichiral superfield, as well as a superfield $\mathcal{S}$ satisfying
$D^2\mathcal{S}=0$ or $\bar{D}^2\mathcal{S}=0$.  For a linear superfield it
also becomes zero, despite the $q+\bar{q}-2$ in the denominator, due to the
``double'' zero from $(q-1)(\bar{q}-1)$ in the numerator. In the
$(\Delta,R)$-representation the $R\to0$ limit produces a $(\Delta-2)^2$ in
the numerator which cancels the $\Delta-2$ in the denominator and thus
leads to zero when $\Delta\to2$.

For a spin-one superconformal operator $\CO^\mu$ a similar treatment,
starting from
\eqn{\langle\CO^\mu(z_1){\bar{\CO}}^\nu(z_2)\rangle =\frac12
C_{\CO^\mu}\frac{\Tr(\bar{\sigma}^\mu\text{x}_{1\bar{2}}
\bar{\sigma}^\nu\text{x}_{2\bar{1}})}
{\osbx{\bar{1}2}{2\bar{q}+1}\osbx{\bar{2}1}{2q+1}},}[]
shows that
\eqn{\langle[Q\sigma_\mu\bar{Q}O^\mu]_p(x)[Q\sigma_\nu\bar{Q}O^\nu]_p
^\dagger(0)\rangle=2^5C_{\CO^\mu}\frac{(2q-3)(2\bar{q}-3)(q+\bar{q}-2)}
{(q+\bar{q}-3)}\frac{1}{x^{2(q+\bar{q}+1)}},}[QQbOmuTP]
where
\eqn{[Q\sigma_{\mu}\bar{Q}O^{\mu}]_{p}=Q\sigma_{\mu}\bar{Q}O^{\mu}
+4\frac{\bar{q}-\frac{3}{2}}{q+\bar{q}-3}P_{\mu}O^{\mu}.}[QQbO31p]
The two-point function coefficient in \QQbOmuTP correctly vanishes if
$\CO^\mu$ is the supercurrent ($D^\alpha
\mathcal{J}_{\alpha\dot{\alpha}}={\bar D}^{\dot{\alpha}}
\mathcal{J}_{\alpha\dot{\alpha}}=0$, $q=\bar{q}=\tfrac32$), in which case
in \QQbOmuTP we see the vanishing of the trace of the stress-energy tensor.
Of course, the condition $D^\alpha \CO_{\alpha\dot{\alpha}}=0$
(respectively, ${\bar{D}}^{\dot\alpha}\CO_{\alpha\dot{\alpha}}=0$) is
enough to shorten a multiplet, and in that case \QQbOmuTP is also zero
since superconformal symmetry requires $q=\tfrac32$ (respectively,
$\bar{q}=\tfrac32$).

Finally, let us mention that in the $R\to0$ limit our expressions \VecTP,
\QsqQbsqTP, and \QQbOmuTP agree with the zero-spin limit of expressions
(3.34), (3.39) and (3.35) of~\cite{Poland:2010wg} respectively.

\newsec{From three-point functions to OPEs}[technicalThreeP]
It is well known that the three-point function of three operators contains
the same information as the OPE of two of them with the third one. Indeed,
a three-point function of the form $\langle\CO_1\CO_2\bar{\CO}_3\rangle$
has to reproduce the two-point function $\langle\CO_3\bar{\CO}_3\rangle$ if
the OPE $\CO_1\times\CO_2$ is used.  In this appendix we will work out this
correspondence explicitly for the $[Q^2\bar{Q}^2O_{3,0}]_p$ contribution to
\muOPE.

We will thus concentrate on the OPE $Q^\alpha O_u(x)\times Q_\alpha
O_d(0)$, which we will recover from the three-point function
\ChiralChiralThreePF.  There are multiple structures for the
${\bar{t}}^{\bar{I}}$ in \ChiralChiralThreePF, but here we will only work
out the term $[Q^2\bar{Q}^2 O_{3,0}]_p$ in \muOPE, which arises from
\ref{SolThree} for $\ell=0$. The coefficient we will compute is
$c_{\mu;4}^i$ in \muOPE.  Other terms in this and other OPEs in section
\ref{computations} can be obtained similarly.

To start, we have to perform the $\theta_{1,2,3}$ and
$\bar{\theta}_{1,2,3}$ expansion of the superconformal three-point function
$\langle\CO_u(z_1)\CO_d(z_2)\CO_{3,0}(z_3)\rangle$ and go to order
$\theta_1\theta_2\,\theta_3^2\,\bar{\theta}_3^{\hspace{0.5pt}2}$. In
practice this a very lengthy computation, but straightforward enough to be
coded in \emph{Mathematica}. As we can see from \SuperFieldExp, the result
of this computation is the combination
\eqna{\frac{1}{2^5}\langle Q^\alpha O_u(x_1)Q_\alpha O_d(x_2)
[Q^2\bar{Q}^2O_{3,0}]_p^\dagger(x_3)\rangle
&-\frac{c_4}{2^5}\langle Q^\alpha O_u(x_1)Q_\alpha
O_d(x_2)(P^2O_{3,0})^\dagger(x_3)\rangle\\
&+\frac{c_5}{2^5}\langle Q^\alpha O_u(x_1)Q_\alpha O_d(x_2)
(P_\mu[Q\sigma^\mu\bar{Q}O_{3,0}]_{p})^\dagger(x_3)\rangle}[CompRes]
of three-point functions involving both primaries and descendants.

Now, in order to compute the Wilson coefficient of
$[Q^2\bar{Q}^2O_{3,0}]_p$ in the OPE $Q^\alpha O_u\times Q_\alpha O_d$, we
need to substitute
\eqn{Q^\alpha O_u(x_1)Q_\alpha O_d(x_2)\sim w(x_{12}^2)
[Q^2\bar{Q}^2O_{3,0}]_p(x_2)}[OPEmuprim]
in \CompRes, and compute $w$ using the known result \QsqQbsqTP for the
resulting two-point function. Obviously, the last two terms in \CompRes
complicate the computation. Indeed, there are other contributions to the
OPE $Q^\alpha O_u\times Q_\alpha O_d$, besides the one in \OPEmuprim, which
also need to be evaluated since they result in nonzero two-point functions
with the conformal descendants $(P^2 O_{3,0})^\dagger$ and
$(P_\mu[Q\sigma^\mu\bar{Q}O_{3,0}]_p)^\dagger$. Since the
$\theta$-expansion of the superconformal three-point function yields
\CompRes, the Wilson coefficients of these extra contributions are
necessary in order to obtain $w(x_{12}^2)$.

Therefore, we need to compute the Wilson coefficients $w_{1,\ldots,4}$ in
\eqna{Q^\alpha O_u(x_1)Q_\alpha O_d(x_2)&\sim
w_1(x_{12}^2)\frac{x_{12}^\mu x_{12}^\nu}{x_{12}^2}
  i^2\partial_\mu\partial_\nu O_{3,0}
+w_2(x_{12}^2)i^2\partial^2 O_{3,0}\\
&\quad+w_3(x_{12}^2)\frac{x_{12}^\mu x_{12}^\nu}{x_{12}^2}
i\partial_\mu[Q\sigma_\nu\bar{Q}O_{3,0}]_p
+w_4(x_{12}^2)i\partial_\mu[Q\sigma^\mu\bar{Q}O_{3,0}]_p,}[]
which in turn require orders $\theta_1\theta_2$ and
$\theta_1\theta_2\,\theta_3\bar{\theta}_3$ of the three-point function
\ChiralChiralThreePF for their evaluation. This is a tedious computation,
but in the end we find that $w(x_{12}^2)=\check{w}
x_{12}^{\Delta_{\CO_{3,0}}\!-\Delta_{\CO_u}- \Delta_{\CO_d}+1}$ with
\eqn{\check{w}=\frac{1}{2^4}\frac{\lambda_{\CO_u\CO_d\CO_{3,0}}}
{C_{\CO_{3,0}}}\frac{(\Delta_{\CO_u}-\Delta_{\CO_d}-\Delta_{\CO_{3,0}}-1)
(\Delta_{\CO_u}-\Delta_{\CO_d}+\Delta_{\CO_{3,0}}+1)}{\Delta_{\CO_{3,0}}
(\Delta_{\CO_{3,0}}+1)(\Delta_{\CO_u}+\Delta_{\CO_d}-\Delta_{\CO_{3,0}}-
1)(\Delta_{\CO_u}+\Delta_{\CO_d}-\Delta_{\CO_{3,0}}-3)},}[]
where $C_{\CO_{3,0}}$ is the coefficient in the superconformal two-point
function $\langle\CO_{3,0}(z_1)\bar{\CO}_{3,0}(z_2)\rangle$, and
$\lambda_{\CO_u\CO_d\CO_{3,0}}$ the coefficient in the superconformal
three-point function $\langle\CO_u(z_1)\CO_d(z_2)
\bar{\CO}_{3,0}(z_3)\rangle$.  As expected, our answer for $w$ is symmetric
under $\CO_u\leftrightarrow\CO_d$.

\end{appendices}

\bibliography{HiggsOPE}

\end{document}